\documentclass[12pt,onecolumn,draftcls]{IEEEtran}

\ifCLASSINFOpdf
\else
\fi
\usepackage{graphicx}
\usepackage{epstopdf}
\usepackage{stfloats}
\usepackage{bm}
\usepackage[cmex10]{amsmath}
\usepackage{algorithmic}
\usepackage{algorithm}
\usepackage{array}

\usepackage{mdwmath}
\usepackage{mdwtab}
\usepackage{multirow}

\usepackage{amsfonts}

\usepackage{amssymb}

\usepackage{color}

\begin{document}
%
\title{User-Centric Distributed Antenna Transmission: Secure Precoding and Antenna Selection with Interference Exploitation}

\author{Zhongxiang Wei,~\IEEEmembership{Member,~IEEE,}
            Christos Masouros,~\IEEEmembership{Senior Member,~IEEE}
\thanks{Zhongxiang Wei and Christos Masouros are with the department of Electronic and Electrical Engineering at the University College London, London, UK. Email: \{zhongxiang.wei, c.masouros\}@ucl.ac.uk}
\thanks{This work was supported by the Engineering and Physical Sciences Research Council, UK, under project EP/R007934/1.} 

}

\maketitle

\begin{abstract}
We address physical layer security in distributed antenna (DA) systems, where eavesdroppers (Eves) can intercept the information transmitted for the intended receiver (IR). 
To realize a user-centric, power-efficient and physical layer security-addressing system, we aim at minimizing total power consumption by jointly designing DA selection and secure precoding. Different from the conventional artificial noise (AN)-aided secure transmission, where AN is treated as an undesired element for the IR, we design AN such that it is constructive to the IR while keeping destructive to the Eves. 
Importantly, we investigate two practical scenarios, where the IR and Eves' channel state information (CSI) is imperfectly obtained or the Eves' CSI is completely unknown. 
To handle the CSI uncertainties, we solve the problems in probabilistic and deterministic robust optimization respectively, both satisfying the IR' signal-to-interference-and-ratio (SINR) requirement by use of constructive AN and addressing security against the Eves.
Simulation results demonstrate our algorithms consume much less power compared to the centralized antenna (CA) systems with/without antenna selection, as well as the DA systems with conventional AN processing. 
Last but not least, by the proposed algorithms, the activation of DAs closely relates to users' locations and quality-of-service (QoS) requirements, featuring a user-centric and on-demand structure.

\end{abstract}


%
\IEEEpeerreviewmaketitle

\section{Introduction}

Wireless Communications for the future Internet of Things (IoT) and Industry 4.0 are required to provide power-efficient transmission together with high security level \cite{Lin2017A}. 
In the last decade, centralized multiple-input multiple-output (MIMO) has been considered as a potential technique due to its high throughput \cite{Xie2017A} and additional spatial diversity for enhancing physical layer security \cite{Ng2015Secure}. However, it requires extremely high power consumption caused by the fully activated antennas, and centralized MIMO often suffers from an equal level of path loss (PL) from the antenna array to one user caused by the co-located antenna (CA) deployment \cite{Duarte2014Design}. 
Besides, edge users in CA deployment may not be well served due to the severe propagation attenuation, otherwise significant transmission power is required for compensating the propagation loss. To create a user-centric and power efficient structure necessary for the IoT and Industry 4.0, distributed antenna (DA) systems have attracted much attention \cite{Wei2018Energy}. By geographically distributing the antennas and hence placing them closer to users, DA systems can reduce the PL impact and obtain blockage-free effect and also facilitate an on-demand network structure by activating those DAs contributing the most.
The concepts of user-centric DA are particularly suited for communications in industrial environments, where antennas can be carefully planned and distributed in the ceilings of large factories to effectively extend network coverage without crucial increment of power consumption, and are therefore a key contender for industry IoT and Industry 4.0 deployments. 
It was pointed out in the FP7 EARTH project \cite{Earth2010D} that for a small-scale communication node (such as DA, femto or pico node), the power consumption is dominated by the power amplifier (PA) and circuit power. Generally, PA power consumption is closely related to the transmission power and drain efficiency at transmitters. On the other hand, circuit power consumption contains multiple power consuming components. 
The power consumption of an active DA mainly comes from digital/analog converter, analog/digital, optical/electrical converter, up-converter, filter, synthesizer, mixers, etc \cite{Joung2014Energy}, while the power consumption of one DA can be significantly reduced by switching it off \cite{Bolla2011Energy}.

Nevertheless, it should be noticed that with the advantages in DA systems, due to the proximity to transmitting antennas, it is also easier for potential eavesdroppers (Eves) to obtain the signal transmitted to the intended receiver (IR), and physical layer security issue in DA systems becomes more challenging. In the past decades, physical layer security has been extensively investigated as a complement to secure wireless communications, where artificial noise (AN) is generated at transmitter to jam potential Eves.
When the Eves' channel state information (CSI) is unknown at the transmitter, isotropic AN could be generated into the null space of the IR's channel \cite{Goel2008Guaranteeing}. When the Eves' CSI is known at the transmitter, AN could be injected to the direction of Eves in a spatial manner, which is more efficient than the isotropic transmission \cite{Li2013Spatially}. 
Regardless of isotropic or spatial AN transmission, AN is treated as an undesired element at the IR and its leakage effect needs to be minimized by precoding design \cite{Mukherjee2011Robust}. Based on AN-aided secure transmission, secrecy rate \cite{Zhou2010Secure} \cite{Wang2018Cooperative} \cite{Li2018Artificial}, outage probability \cite{Lei2016Secrecy} and power consumption \cite{Ng2015Secure} have been addressed recently. 
Zhou el in \cite{Zhou2010Secure} derived an analytical closed-form expression of achievable secrecy rate over fading channel.  Wang el in \cite{Wang2018Cooperative} extended the secrecy rate into a two-hop relay system, where the relay is powered by energy harvesting. The authors in \cite{Li2018Artificial} further extended secrecy rate into a two-way relay system. In \cite{Lei2016Secrecy}, the authors investigated the outage probability of the IR in cognitive systems. 
On the specific topic of secure transmission in DA systems, the authors in \cite{Guo2016Security} investigated SINR maximization problem for DA systems, where AN is generated by DAs to interfere the Eves. The authors in \cite{Wang2016Artificial} maximized the ergodic secrecy rate in DA systems, where AN and signal are jointly designed and transmitted at each DA. In \cite{Ng2015Secure}, the power minimization problem was demonstrated for DA systems, where the potential Eves are considered as idle IRs and scavenge energy from AN. 


The aforementioned research, however, treated AN as a catastrophic element and mitigated the effect of AN at the IR as much as possible. If AN can be carefully designed at the transmitter, it may be beneficial to the IR in terms of improving signal-to-interference-plus-noise ratio (SINR) based on the concept of constructive interference (CI). The concept of CI was firstly introduced by \cite{Christos2007A} in CDMA system. Then a rotated zero-forcing (ZF) precoding scheme was proposed in \cite{Christos2007A} that partially CI can be utilized while the destructive interference should be suppressed, and \cite{Christos2011Correlation} further proposed that all the interference can be constructive by designing precoding in symbol level. 
The concept of CI was adapted to beamforming optimization in \cite{Masouros2014Vector} in the context of symbol scaling, and subsequently in \cite{Alodeh2015Constructive} and \cite{Masouros2015Exploiting}. 
Recently, the concept of CI was applied into cognitive radio \cite{Law2017Transmit}, large-scale multi-input multi-output  (MIMO) \cite{Amadori2017Large}, multiuser multi-input single-output (MISO) \cite{Li2018Interference}, wireless power transfer \cite{Timotheou2016Exploiting} and constant envelop systems \cite{Amadori2017Constant}, 
Based on the aforementioned work in CI design, the authors in \cite{Khandaker2018Constructive} proposed a scheme to utilize AN in a CA system. However, the fully activated antennas in \cite{Khandaker2018Constructive} lead to enormous power consumption.

Motivated by the aforementioned issues, in this paper, we present joint design of DA selection and secure precoding to minimize total power consumption, subjected to physical layer security constraints. Importantly, we consider two practical scenarios and solve the problems from two different prospectives of robust optimizations, namely probabilistic and deterministic robust optimizations. Our contributions are summarized in the following:

\begin{enumerate}[]
	\item We investigate the power minimization problem under the IR's QoS requirement and physical layer security constraints against Eves. DA selection vector and precoding are jointly designed to fully utilize the additional degrees of freedom in antennas' activation/deactivation and beneficial effect of AN, which is shown to significantly reduce the total power consumption yet maintaining the IR's SINR and security constraints against the Eves.
	
	\item We exploit joint DA selection and robust precoding in two practical scenarios: first, when the IR and Eves' CSI is imperfectly obtained, and second, when the Eves' CSI is completely unknown. Then we investigate the total power minimization problems for the two scenarios in probabilistic and deterministic manners, respectively. In the first scenario, the IR's SINR and the security against the Eves are issued by chance constrained formulations from the prospective of probabilistic robust optimization, while the IR's SINR and security against the Eves are guaranteed with all the CSI uncertainties from the prospective of deterministic robust optimization. On the other hand, when the Eves's CSI in completely unknown at the transmitter side in the second scenario, the IR's SINR requirement is addressed by the probabilistic or deterministic robust optimization, while the security towards the Eves is addressed by confining a minimum power level of AN.
	
	\item Four corresponding low-complexity algorithms are proposed to minimize total power consumption for the two scenarios in terms of probabilistic and deterministic manners, and at the same time AN is kept constructive to the IR whereas destructive to the Eves. Complexities of the algorithms are analytically demonstrated.
	
    \item A user-centric network structure is demonstrated by the proposed schemes, compared to the CA counterpart. Explicitly, the working status of each DA is flexibly determined by the IR and Eves' positions to provide on-demand services: the DAs close to users have higher probabilities of activation, while the DAs far from users have higher probabilities of being idle for saving power. Furthermore, the power consumption of the DA deployment remains low regardless of the users' positions and power-efficient transmission is always featured, while the power consumption of the CA deployment demonstrates a significant increment when the users move to edge area.
\end{enumerate}

\textit{Notations}:
Matrices and vectors are represented by boldface capital and lower case letters, respectively. $\vert\cdot \vert$ denotes the absolute value of a complex scalar. $\vert\vert\cdot \vert\vert$ denotes the Euclidean vector norm. $\bm{A}^H$ $\bm{A}^T$ and Tr$(\bm{A})$ denote the Hermitian transpose, transpose and trace of matrix $\bm{A}$. Rank($\bm{A}$) denote the rank of matrix $\bm{A}$. diag ($\bm{A}$) returns a diagonal matrix with diagonal elements from matrix $\bm{A}$ and diag ($\bm{a}$) stacks the elements of vector $\bm{a}$ into a diagonal matrix. $\bm{A}\succeq 0$ means $\bm{A}$ is a positive semi-definite matrix. Superscript $n$ denotes the $n$-th DA' index or the $n$-th element of a vector. $\parallel \cdot \parallel_p$ means the p-norm of a vector or a matrix. $\bm{I}_n$ means a $n$-by-$n$ identity matrix. $\Re$ and $\Im$ represent the real part and imaginary parts, respectively.  $ \mathbb{C}^{N\times M}$ and $ \mathbb{H}^{N\times M}$  denote sets of all $N\times M$ matrices and Hermitian matrices with complex entries.

\section{System Model and Constructive Interference  }
In this section, system model is introduced in II-A and the concept of CI is briefly discussed in II-B.

\subsection{System Model}
We consider a DA system at downlink transmission, which is depicted in Fig. 1. All DA ports are connected to the central unit through a noise-free wired front-haul for cooperative communications.
Central unit is equipped with $N$ DAs transmitting confidential message to the IR in the presence of $K$ possible Eves. 
The IR and Eves are all equipped with single antenna for simplicity. 
CSI is obtained by channel estimation in the training phase, based on channel reciprocity as in \cite{Wei2018Energy} \cite{Sun2016Multi}. Without loss of generality, we assume that all the DAs share the same drain efficiency and their PAs work in the linear region. Per-DA power constraint is applied, which is essentially different from joint power constraint in CA systems.

\begin{figure}
	\centering
	\includegraphics[width=3.1 in]{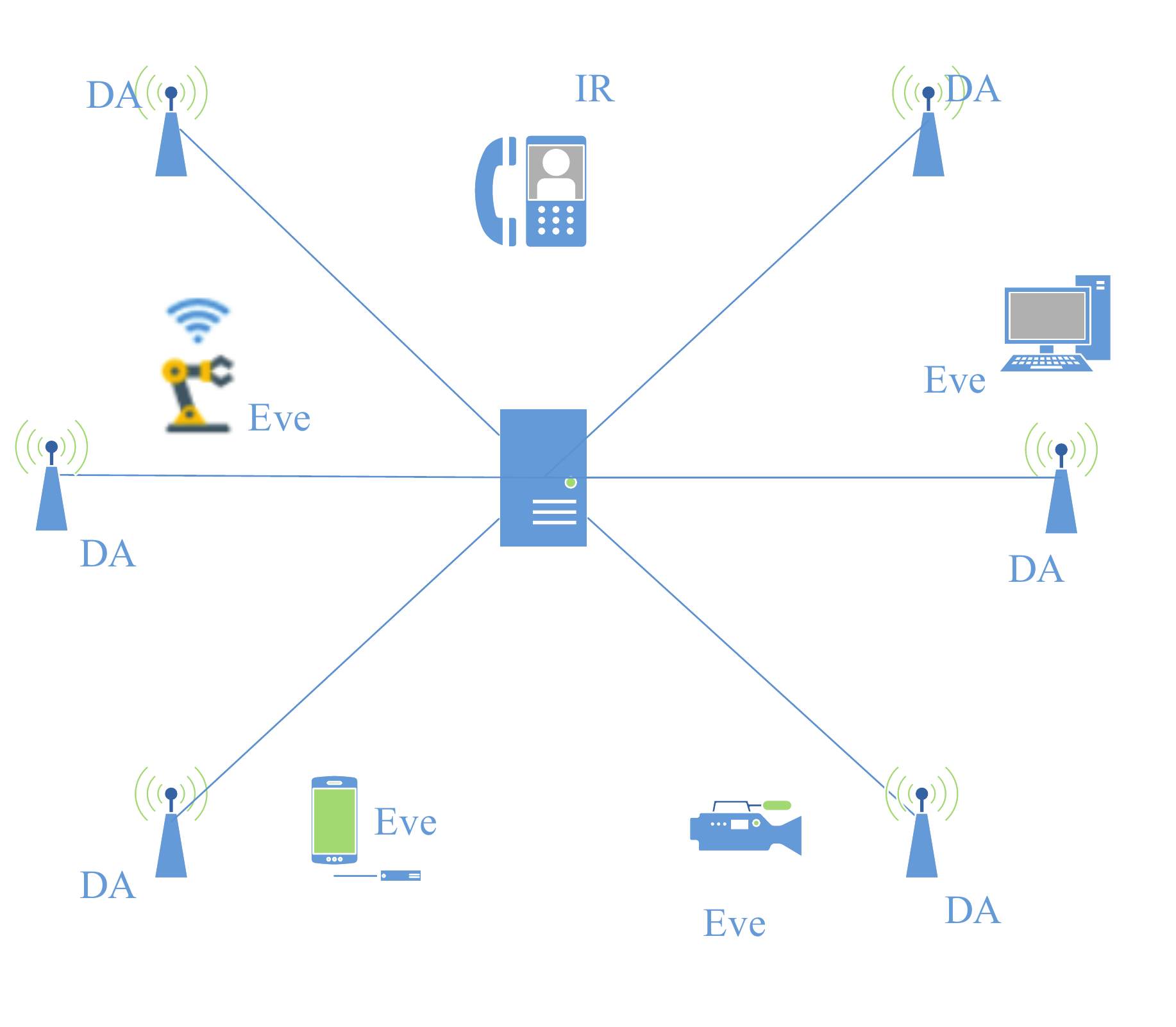}
	\caption{ Illustration of system model, where DAs are geographically positioned.  }
	\label{fig:system20180711}
\end{figure}

\subsection{Constructive Interference}
CI is the interference pushes to received signals aways from the detection threshold \cite{Christos2011Correlation}. The increased distance to the detection threshold can effectively improve the receiving performance. Denote $\bm{w}\in \mathbb{C}^{N\times1}$ and $\bm{z}\in \mathbb{C}^{N\times1}$ as precoding and AN at the transmitter side. Denote $x_d=de^{j\phi_d}$ as the information-bearing symbol transmitted for the IR. The received signal at the IR and the $k$-th Eve can be calculated as

\begin{small}
\begin{equation}
\begin{split}
& y_d= \bm{h}_{d}^T (\bm{w}x_d+\bm{z})+ n_d,~y_k= \bm{h}_{k}^T (\bm{w}x_d+\bm{z})+ n_k,
\label{eq:Recevied signal}
\end{split}
\end{equation}
\end{small}%
where $\bm{h}_d \in \mathbb{C}^{N\times1}$ denotes the channel conditions between the DAs and the IR. $\bm{h}_k \in \mathbb{C}^{N\times1}$  denotes the channel conditions between the DAs and the $k$-th Eve. $n_d \sim \mathcal{CN}(0,\sigma_n^2)$ and $n_k\sim \mathcal{CN}(0,\sigma_n^2)$ denote the Additive white Gaussian noises (AWGN) at the IR and the $k$-th Eve, respectively.
Conventionally, the received SINR at the IR and $k$-th Eve are denoted as

\begin{small}
\begin{equation}
\begin{split}
&  \Gamma_{d}=\dfrac{ | \bm{h}_{d}^T\bm{w}|^2 }{ \sigma_n^2+|\bm{h}_{d}^T\bm{z}|^2     },~\Gamma_{k}=\dfrac{ | \bm{h}_{k}^T\bm{w}|^2 }{ \sigma_n^2 +|\bm{h}_{k}^T\bm{z}|^2      },
\label{eq:Con d and eve}
\end{split}
\end{equation}
\end{small}%
where $\Gamma_{d}$ and $\Gamma_{k}$ denote the IR and the $k$-th Eve's SINR, respectively. 
It can be seen from (\ref{eq:Con d and eve}) that AN is treated as an undesired element at the IR. 
By contrast, the principle of constructive AN is to rotate the phase of the AN at transmitter and to align it with the desired signal at the IR. Since the transmitted signal can be also written as $ (\bm{w}+\bm{z}e^{-j\phi_d})x_d$, according to the principle of CI, the requirement of generating constructive AN can be given as

\begin{small}
\begin{equation}
\begin{split}
& \angle ( \bm{h}_{d}^T(\bm{w}+\bm{z}e^{-j\phi_d})     )=\angle(x_d),  \frac{ \Re \{     \bm{h}_{d}^T(\bm{w}+\bm{z}e^{-j\phi_d})      \}^2 }{\sigma_n^2} \geq \Gamma_d,  
\label{eq:strict CI}
\end{split}
\end{equation}
\end{small}%
where $\angle$ represents the angle. The angular requirement in  (\ref{eq:strict CI}) is referred as strict phase rotation CI design \cite{Li2018Interference}. By exploiting the geometric interpretation in Fig. 2, non-strict phase rotation can be adopted to relax the strict angular requirement without loss of optimality  \cite{Khandaker2018Constructive}, which is mathematically given by

\begin{small}
\begin{figure}
	\centering
	\includegraphics[width=3.5 in]{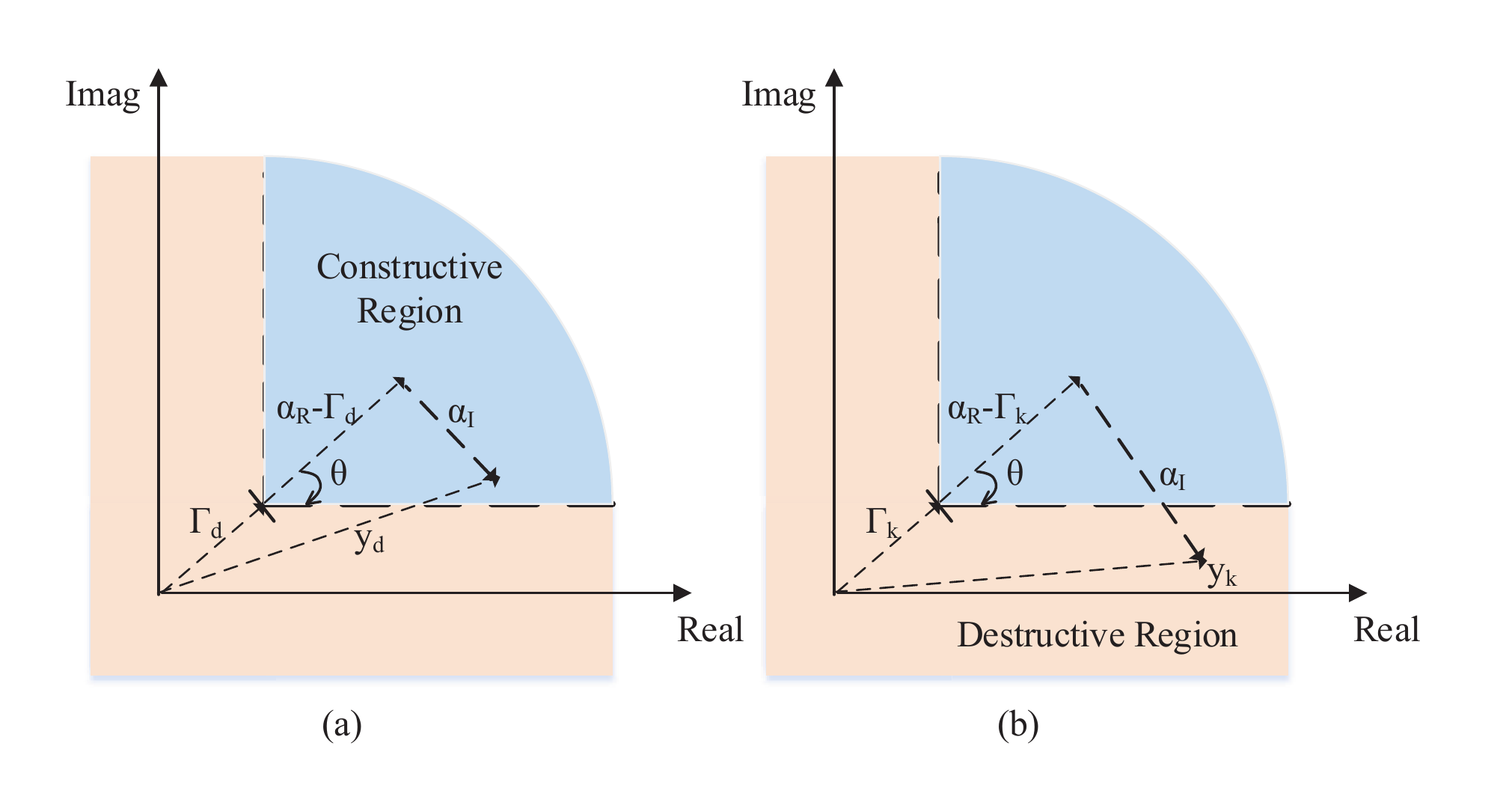}
	\caption{Constructive and destructive AN for the IR and Eves with QPSK. (a) Constructive AN pushes the IR's received symbols towards the constructive region. (b) Destructive AN pushes the Eves' received symbols towards the destructive region. }
	\label{fig:geometric interpretation}
\end{figure}
\end{small}%

\begin{small} 
\begin{equation}
\begin{split}
 |\Im \{  \bm{h}_{d}^T\bm{w}+\bm{h}_{d}^T\bm{z}e^{-j\phi_{d}}  \}|  \leq (\Re \{   \bm{h}_{d}^T\bm{w}+\bm{h}_{d}^T\bm{z}e^{-j\phi_{d}} \}-\sigma\sqrt{\Gamma_{d}}  )            \cdot \mathrm{tan}\theta,
\label{eq:relaxed IR}
\end{split}
\end{equation}
\end{small}%
where $\theta=\pi /M$ and $M$ is constellation size \cite{Masouros2015Exploiting}. It can be seen from (\ref{eq:relaxed IR}) that the AN becomes a beneficial element to the IR and the IR' SINR $\Gamma_d$ is also embedded. Since the AN contributes to the useful signal power, the received SINR of the IR becomes into 
\begin{small}
\begin{equation}
\begin{split}
&  \gamma_{d}=\dfrac{ | \bm{h}_{d}^T(\bm{w}+  \bm{z}e^{-j\phi_{d}}  )  |^2 }{ \sigma_n^2    },
\label{eq:CI SINR d}
\end{split}
\end{equation}
\end{small}%
where the received SINR at the IR is improved by utilizing AN and thus transmission power can be efficiently reduced to achieve a target SINR requirement.


\section{Power Efficient DA Selection and Secure Precoding with Imperfect CSI}

In Section III, we  investigate the power efficient design with imperfect CSI of all the nodes, where the channels of all the nodes are given as $\bm{h}_d=\hat{\bm{h}_d}+\bm{e}_d$ and $\bm{h}_k=\hat{\bm{h}_k}+\bm{e}_k, \forall k \in K$. $\hat{\bm{h}_d} \in \mathbb{C}^{N\times1}$ denotes the estimated channel between the DAs and the IR with estimation error $\bm{e}_d \in \mathbb{C}^{N\times1}$. $\hat{\bm{h}_k} \in \mathbb{C}^{N\times1}$ denotes the estimated channel between the DAs and the $k$-th Eve with estimation error $\bm{e}_k \in \mathbb{C}^{N\times1}$. We assume that the channel estimation error of the IR and the $k$-th Eve follows normal distribution $\bm{e}_d \sim \mathcal{CN}\{0,\sigma_d^2\}$ and $\bm{e}_k \sim \mathcal{CN}\{0,\sigma_k^2\}, \forall k \in K$. To fully exploit the power efficient design with the uncertainties (CSI error), we handle the optimization problem in probabilistic and deterministic manners, respectively.


\subsection{Probabilistic Robust Optimization}

\subsubsection{Problem Formulation}
Define precoding vector $\bm{w} \in \mathbb{C}^{N\times1}$, whose $n$-th element $w_n$ represents the precoding weight at the $n$-th DA. Define AN vector $\bm{z}\in \mathbb{C}^{N\times1}$, whose element $z_n$ represents the AN generated at the $n$-th DA. Define DA selection vector $\bm{t}$, whose element $t_n=\{0,1\}$ means the $n$-th DA is deactivated or activated, respectively. Taking advantage of CI, AN is properly rotated such that it contributes to the received signal power at the IR while remaining destructive to the Eves. 
To minimize the total power consumption, we jointly optimize precoding $\bm{w}$, AN $\bm{z}$ and DA selection vector $\bm{t}$. Accordingly, the problem is formulated as

\begin{small}
\begin{equation}
\begin{split}
&P1~ (\mathrm{imperfect-prob}): \operatorname*{argmin}\limits_{\bm{w},\bm{z},\bm{t}}   \dfrac{||  \bm{w}+\bm{z}e^{-j\phi_{d}}  ||^2}{\alpha}    + \sum_{n=1}^{N}\big(t_np_{on}+(1-t_n)p_{off} \big),\\
&~~~~~\mathrm{s.t}~(C1):0\leq |w_n+z_ne^{-j\phi_{d}}|^2\leq t_np_{DA}, \forall n\in N,~(C2):t_n=\{0,1\}  ,\forall n\in N,\\
& ~~~~~~~~~(C3): \mathrm{Pr}\{\ \Gamma_{d} \geq \overline{ \Gamma_{d} } |  \bm{e}_{d}   \} \geq \eta_{d},~(C4): \mathrm{Pr}\{\ \Gamma_{k} \leq \overline{ \Gamma_{k} } |  \bm{e}_{k}   \} \geq \eta_{k}, \forall k \in K,
\label{eq:probabilistic_P1}
\end{split}
\end{equation}
\end{small}%
where $\alpha$ is the drain efficiency of the DAs. $p_{DA}$ denotes the maximum available transmission power at each DA.
$p_{on}$ and $p_{off}$ represent the power consumption of each activated/deactivated DA, respectively.
$\overline{ \Gamma_{d} } $ and $\overline{ \Gamma_{k} } $ are the SINR requirement for the IR and physical layer security against the $k$-th Eve. $\eta_{d}$ and $\eta_{k}$ denote the probabilistic thresholds for the IR and the $k$-th Eve, respectively. Evidently, $(C1)$ imposes individual transmission power constraint at each DA, which is different from the joint transmission power constraint in CA systems. $(C2)$ constrains the selection vector to binary (on/off) elements. 
$(C3)$ and $(C4)$ guarantee that the SINR constraints at the IR and the $k$-th Eve with probabilities greater than $\eta_d$ and lower than $\eta_k, \forall k \in K$.

\subsubsection{Solution to the Problem}
To solve the problem, we first need to handle the probabilistic constraints $(C3)$ and $(C4)$.  Defining $ \bm{u}=\bm{w}+\bm{z}e^{-j\phi_{d}}$, under the provision of CI the constraint $(C3)$ equals to

\begin{small}
\begin{equation}
\begin{split}
& (C3): \mathrm{Pr}\{\ \Gamma_{d} \geq \overline{ \Gamma_{d} } |  \bm{e}_{d}   \} \geq \eta_{d} \overset{(\ref{eq:relaxed IR})}{\Rightarrow} \mathrm{Pr}\{ |\Im \{ \bm{h}_{d} ^T \bm{u} \} | \leq   \big( \Re \{ \bm{h}_{d}^T  \bm{u} \}  -\sigma_n \sqrt{ \overline {\Gamma_{d}} } \big)       \mathrm{tan}\theta | \bm{e}_{d} \} \geq \eta_{d}\Rightarrow\\
& ~~~~~~~~~\mathrm{Pr}\{ |\Im \{ ( \hat{\bm{h}_{d}} +\bm{ e}_{d} )^T \bm{u} \} | \leq   \big( \Re \{ ( \hat{\bm{h}_{d}}+\bm{e}_{d})^T\bm{u}   \}  -\sigma_n \sqrt{ \overline {\Gamma_{d}} } \big)       \mathrm{tan}\theta | \bm{e}_{d} \} \geq \eta_{d}.
\label{eq:equivalent IR}
\end{split}
\end{equation}
\end{small}%
	
Decomposing the real part and imaginary parts, (\ref{eq:equivalent IR}) is equivalent to the two constraints in (\ref{eq:equivalent IR2}). 

\begin{small}
\begin{equation}  
\left\{  
\begin{array}{lr}  
\mathrm{Pr}\{ \sigma_n\sqrt{ \overline{  \Gamma_{d}}} \mathrm{tan}\theta+\hat{\bm{ h}_{I,d}} \bm{u}_{R}+ \hat{ \bm{h}_{R,d}} \bm{u}_I+  \bm{e}_{I,d}\bm{u}_{R} + \bm{e}_{R,d} \bm{u}_I  \\
 -(\hat{ \bm{h}_{R,d}} \bm{u}_{R}-\hat{ \bm{h}_{I,d}} \bm{u}_I+ \bm{e}_{R,d} \bm{u}_{R} - \bm{e}_{I,d} \bm{u}_I) \mathrm{tan}\theta  \leq 0| \bm{e}_{d}\} \geq \eta_d, &  \\  
\mathrm{Pr}\{\sigma_n\sqrt{ \overline{  \Gamma_{d}}} \mathrm{tan}\theta-\hat{ \bm{h}_{I,d}} \bm{u}_{R}- \hat{ \bm{h}_{R,d}} \bm{u}_I- \bm{e}_{R,d}\bm{u}_{I} - \bm{e}_{I,d} \bm{u}_R\\
 -(\hat{ \bm{h}_{R,d}} \bm{u}_{R}-  \hat{ \bm{h}_{I,d}} \bm{u}_I+ \bm{e}_{R,d} \bm{u}_{R} -\bm{e}_{I,d} \bm{u}_I) \mathrm{tan}\theta  \leq 0 | \bm{e}_{d}\} \geq \eta_d.
\end{array}  
\right.  
\label{eq:equivalent IR2}
\end{equation} 
\end{small}%

We now handle the first inequality in (\ref{eq:equivalent IR2}), which equals to $\mathrm{Pr}\{ \bm{a_{d,1}}^T[\bm{u}_R;\bm{u}_I] \leq -\sigma_n\sqrt{ \overline{  \Gamma_{d}}}\mathrm{tan}\theta| \bm{e}_{d}\} \\ \geq \eta_d$, where $ \bm{a}_{d,1}=[ \hat{ \bm{h}_{I,d}}-\hat{ \bm{h}_{R,d}}\mathrm{tan}\theta+\bm{e}_{I,d}-\bm{e}_{R,d}\mathrm{tan}\theta;  \hat{\bm{ h}_{R,d}}+\hat{ \bm{h}_{I,d}} \mathrm{tan}\theta+\bm{e}_{R,d}+\bm{e}_{I,d}\mathrm{tan}\theta]$. 
It is easy to obtain that the $N$-dimensional normal distributed vector $\bm{a}_{d,1}$'s expectation is $ \overline{\bm{a}_{d,1}} =[ \hat{ \bm{h}_{I,d}}-\hat{ \bm{h}_{R,d}}\mathrm{tan}\theta,\hat{ \bm{h}_{R,d}}+\hat{ \bm{h}_{I,d}} \mathrm{tan}\theta]$ with covariance matrix $\bm{\Theta}_{d,1}=\mathrm{diag} ( \underbrace{(1+\mathrm{tan}\theta)^2\sigma_e^2,...,(1+\mathrm{tan}\theta)^2\sigma_e^2}_{2N})   $. Hence, we know that 


\begin{small}
\begin{equation}
\begin{split}
& \mathrm{Pr}\{\  \bm{a}_{d,1}^T[\bm{u}_R;\bm{u}_I] \leq -\sigma_n\sqrt{ \overline{  \Gamma_{d}}}\mathrm{tan}\theta |  \bm{e}_{d}   \} \geq \eta_{d}  \Rightarrow\\
&\mathrm{Pr}\{ \frac{\bm{a}_{d,1}^T[\bm{u}_R;\bm{u}_I]- \overline{\bm{a}_{d,1}}^T[\bm{u}_R;\bm{u}_I]}{|| \bm{\Theta}^{\frac{1}{2}}_{d,1} [\bm{u}_R;\bm{u}_I] ||_2}\leq   \frac{ -\sigma_n\sqrt{ \overline{  \Gamma_{d}}}\mathrm{tan}\theta-\overline{\bm{a}_{d,1}}^T[\bm{u}_R;\bm{u}_I]}{|| \bm{\Theta}^{\frac{1}{2}}_{d,1} [\bm{u}_R;\bm{u}_I] ||_2}   |  \bm{e}_{d}   \} \geq \eta_{d} \Rightarrow\\
&  \Phi(  \frac{ -\sigma_n \sqrt{ \overline{  \Gamma_{d}}}\mathrm{tan}\theta-\overline{\bm{a}_{d,1}}^T[\bm{u}_R;\bm{u}_I]}{|| \bm{\Theta}^{\frac{1}{2}}_{d,1} [\bm{u}_R;\bm{u}_I] ||_2}   ) \geq \eta_{d},
\label{eq:equivalent IR5}
\end{split}
\end{equation}
\end{small}%
where $\Phi(x)=\frac{2}{\sqrt{\pi}} \int_{0}^{x}e^{-t^2} \mathrm{d}t$ denotes the cumulative probability function (cdf) of a standard normal distributed variable. Defining $\Phi^{-1}(^.)$ as the inverse function of $\Phi(^.)$, (\ref{eq:equivalent IR5}) can be finally derived into a quadratic constraint 

\begin{small}
\begin{equation}
\begin{split}
\overline{\bm{a}_{d,1}}^T[\bm{u}_R;\bm{u}_I]+\Phi^{-1}(\eta_{d} )|| \bm{\Theta}^{\frac{1}{2}}_{d,1} [\bm{u}_R;\bm{u}_I] ||_2 \leq  -\sigma_n\sqrt{ \overline{  \Gamma_{d}}}\mathrm{tan}\theta,
\label{eq:equivalent IR6}
\end{split}
\end{equation}
\end{small}%

Similarly, the second equation in (\ref{eq:equivalent IR2}) can be given as  

\begin{small}
\begin{equation}
\begin{split}
\overline{\bm{a}_{d,2}}^T [\bm{u}_R,\bm{u}_I]+\Phi^{-1}(\eta_{d} )|| \bm{\Theta}^{\frac{1}{2}}_2 [\bm{u}_R,\bm{u}_I] ||_2 \leq  -\sigma_n\sqrt{ \overline{  \Gamma_{d}}}\mathrm{tan}\theta.
\label{eq:equivalent IR7}
\end{split}
\end{equation}
\end{small}%
where $\overline{\bm{a}_{d,2}} =[- \hat{ \bm{h}_{I,d}}-\hat{ \bm{h}_{R,d}}\mathrm{tan}\theta;-\hat{ \bm{h}_{R,d}}+\hat{ \bm{h}_{I,d}} \mathrm{tan}\theta]$ with covariance matrix calculated as $\bm{\Theta}_{d,2}=\mathrm{diag} ( \underbrace{(1+\mathrm{tan}\theta)^2\sigma_e^2,...,(1+\mathrm{tan}\theta)^2\sigma_e^2}_{2N})  $.  
Hence, the IR's probabilistic SINR constraint in $(C3)$ is equivalent to the following two inequalities

\begin{small}
\begin{equation}  
\left\{  
\begin{array}{lr}  
\overline{\bm{a}_{d,1}}^T [\bm{u}_R,\bm{u}_I]+\Phi^{-1}(\eta_{d} )|| \bm{\Theta}^{\frac{1}{2}}_{d,1} [\bm{u}_R,\bm{u}_I] ||_2 \leq  -\sigma_n\sqrt{ \overline{  \Gamma_{d}}}\mathrm{tan}\theta,&  \\  
\overline{\bm{a}_{d,2}}^T [\bm{u}_R,\bm{u}_I]+\Phi^{-1}(\eta_{d} )|| \bm{\Theta}^{\frac{1}{2}}_{d,2} [\bm{u}_R,\bm{u}_I] ||_2 \leq  -\sigma_n\sqrt{ \overline{  \Gamma_{d}}}\mathrm{tan}\theta,
\end{array}  
\right.  
\label{eq:equivalent IR8}
\end{equation}  
\end{small}%

We now handle the security constraint against the $k$-th Eve in $(C4)$ under the provision of destructive interference. According to the geometric interpretation, confining the $k$-th Eve in the destructive region equals to $|\Im \{  \bm{h}_{k}^T\bm{u}   \}|  \geq \big(\Re \{   \bm{h}_{k}^T\bm{u} \}-\sigma_n\sqrt{\overline{\Gamma_{k}}  }        \big)     \mathrm{tan}\theta$. Hence, the constraint $(C4)$ holds if the two constraints in (\ref{eq:equivalent_eve2}) are simultaneously satisfied




\begin{small}
\begin{equation} 
\begin{split} 
\left\{  
\begin{array}{lr}  
\mathrm{Pr}\{ \bm{a_{k,1}}^T[\bm{u}_R;\bm{u}_I] \leq \sigma_n\sqrt{ \overline{  \Gamma_{k}}}\mathrm{tan}\theta| \bm{e}_{k}\} \geq \eta_k. &  \\  
\mathrm{Pr}\{ \bm{a_{k,2}}^T[\bm{u}_R;\bm{u}_I] \leq \sigma_n\sqrt{ \overline{  \Gamma_{k}}}\mathrm{tan}\theta| \bm{e}_{k}\} \geq \eta_k. 
\end{array}  
\right.  
\label{eq:equivalent_eve2}
\end{split}
\end{equation}  
\end{small}%
where $ \bm{a}_{k,1} =[ \hat{ \bm{h}_{I,k}}+\hat{ \bm{h}_{R,k}}\mathrm{tan}\theta+\bm{e}_{I,k}+\bm{e}_{R,k}\mathrm{tan}\theta;\hat{ \bm{h}_{R,k}}-\hat{ \bm{h}_{I,k}} \mathrm{tan}\theta+\bm{e}_{R,k}-\bm{e}_{I,k}\mathrm{tan}\theta]$ and $ \bm{a}_{k,2} =[ -\hat{ \bm{h}_{I,k}}+\hat{ \bm{h}_{R,k}}\mathrm{tan}\theta-\bm{e}_{I,k}+\bm{e}_{R,k}\mathrm{tan}\theta;-\hat{ \bm{h}_{R,k}}-\hat{ \bm{h}_{I,k}} \mathrm{tan}\theta-\bm{e}_{R,k}-\bm{e}_{I,k}\mathrm{tan}\theta]$.
Then the probabilistic constraint $(C4)$ can be transformed into

\begin{small}
\begin{equation}  
\left\{  
\begin{array}{lr}  
\overline{\bm{a}_{k,1}}^T[\bm{u}_R;\bm{u}_I]+\Phi^{-1}(\eta_{k} )|| \bm{\Theta}^{\frac{1}{2}}_{k,1}[\bm{u}_R;\bm{u}_I] ||_2 \leq  \sigma_n\sqrt{ \overline{  \Gamma_{k}}}\mathrm{tan}\theta , &  \\  
\overline{\bm{a}_{k,2}}^T[\bm{u}_R;\bm{u}_I]+\Phi^{-1}(\eta_{k} )|| \bm{\Theta}^{\frac{1}{2}}_{k,2}[\bm{u}_R;\bm{u}_I] ||_2 \leq  \sigma_n\sqrt{ \overline{  \Gamma_{k}}}\mathrm{tan}\theta,  
\end{array}  
\right.  
\label{eq:equivalent_eve3}
\end{equation}  
\end{small}%
where $ \overline{\bm{a}_{k,1}} =[ \hat{ \bm{h}_{I,k}}+\hat{ \bm{h}_{R,k}}\mathrm{tan}\theta,\hat{ \bm{h}_{R,k}}-\hat{ \bm{h}_{I,k}} \mathrm{tan}\theta]$ and $ \overline{\bm{a}_{k,2}} =[ -\hat{ \bm{h}_{I,k}}+\hat{ \bm{h}_{R,k}}\mathrm{tan}\theta,-\hat{ \bm{h}_{R,k}}-\hat{ \bm{h}_{I,k}} \mathrm{tan}\theta]$ with covariance matrix $ \Theta_{k,1}=\Theta_{k,2}=\mathrm{diag} ( \underbrace{(1+\mathrm{tan}\theta)^2\sigma_e^2,...,(1+\mathrm{tan}\theta)^2\sigma_e^2}_{2N}) $.

After a series of transformation, the probabilistic constraints $(C3)$ and $(C4)$ in P1(imperfect-prob) are transformed into the corresponding quadratic constraints in (\ref{eq:equivalent IR8}) and (\ref{eq:equivalent_eve3}). The optimization problem P1(imperfect-prob) can be re-formulated as

\begin{small}
\begin{equation}
\begin{split}
& P2~ (\mathrm{imperfect-prob}): \operatorname*{argmin}\limits_{\bm{u},t_n, n\in N}   \dfrac{||  \bm{u}  ||^2}{\alpha}    + \sum_{n=1}^{N}\big(t_np_{on}+(1-t_n)p_{off} \big),\\
&~\mathrm{s.t~}(C1):0\leq |u_n|^2\leq t_np_{DA}, \forall n\in N,~(C2):t_n=\{0,1\}  ,\forall n\in N,~(C3):(\ref{eq:equivalent IR8}),~(C4):(\ref{eq:equivalent_eve3}), \forall k \in K.
\label{eq:probabilistic_P2}
\end{split}
\end{equation}
\end{small}%

According to Schur Complements that $ ||\bm{A}\bm{x}+\bm{b}||_2\leq \bm{e}^T\bm{x}+d$  is equivalent to 
$\left[
\begin{smallmatrix}
(\bm{e}^T\bm{x}+d)\bm{I}& \bm{A}\bm{x}+\bm{b}\\
(\bm{A}\bm{x}+\bm{b})^T& \bm{e}^T\bm{x}+d\\
\end{smallmatrix}  
\right]\succeq  \bm{0}$   \cite{Boyd2004Convex}, the constraints (C3) and (C4)  can be further transformed into (\ref{eq:Schur1}) and (\ref{eq:Schur2}), respectively.

\begin{equation}  
\left\{  
\begin{array}{lr}  
 \left[
 \begin{smallmatrix}
\frac{ \big(-\overline{\bm{a}_{d,1}}^T[\bm{u}_R,\bm{u}_I]-\sigma_n\sqrt{ \overline{  \Gamma_{d}}}\mathrm{tan}\theta\big) \bm{I}}{\Phi^{-1}(\eta_{d}) }& \bm{\Theta}^{\frac{1}{2}}_{d,1}[\bm{u}_R,\bm{u}_I] \\
  (\bm{\Theta}^{\frac{1}{2}}_{d,1}[\bm{u}_R,\bm{u}_I]) ^T& \frac{\big(-\overline{\bm{a}_{d,1}}^T[\bm{u}_R,\bm{u}_I]-\sigma_n\sqrt{ \overline{  \Gamma_{d}}}\mathrm{tan}\theta\big)}{\Phi^{-1}(\eta_{d}) } \\
 \end{smallmatrix}
 \right]  \succeq \bm{0}, &  \\  
 \\
\left[
\begin{smallmatrix}
\frac{ \big(-\overline{\bm{a}_{d,2}}^T[\bm{u}_R,\bm{u}_I]-\sigma_n\sqrt{ \overline{  \Gamma_{d}}}\mathrm{tan}\theta\big) \bm{I}}{\Phi^{-1}(\eta_{d}) }&  \bm{\Theta}^{\frac{1}{2}}_{d,2}[\bm{u}_R,\bm{u}_I] \\
 (\bm{\Theta}^{\frac{1}{2}}_{d,2}[\bm{u}_R,\bm{u}_I] )^T& \frac{\big(-\overline{\bm{a}_{d,2}}^T[\bm{u}_R,\bm{u}_I]-\sigma_n\sqrt{ \overline{  \Gamma_{d}}}\mathrm{tan}\theta\big)}{\Phi^{-1}(\eta_{d}) } \\
\end{smallmatrix}
\right] \succeq \bm{0}
\end{array}  
\right.  
\label{eq:Schur1}
\end{equation}

\begin{equation}  
\left\{  
\begin{array}{lr}  
\left[
\begin{smallmatrix}
\frac{ \big(-\overline{\bm{a}_{k,1}}^T[\bm{u}_R,\bm{u}_I]+\sigma_n\sqrt{ \overline{  \Gamma_{k}}}\mathrm{tan}\theta\big) \bm{I}}{\Phi^{-1}(\eta_{k}) }&  \bm{\Theta}^{\frac{1}{2}}_{k,1}[\bm{u}_R,\bm{u}_I] \\
(\bm{\Theta}^{\frac{1}{2}}_{k,1}[\bm{u}_R,\bm{u}_I])^T& \frac{\big(\overline{-\bm{a}_{k,1}}^T[\bm{u}_R,\bm{u}_I]+\sigma_n\sqrt{ \overline{  \Gamma_{k}}}\mathrm{tan}\theta\big)}{\Phi^{-1}(\eta_{k}) } \\
\end{smallmatrix}
\right] \succeq \bm{0}, &  \\  
\\
\left[
\begin{smallmatrix}
\frac{ \big(\overline{-\bm{a}_{k,2}}^T[\bm{u}_R,\bm{u}_I]+\sigma_n\sqrt{ \overline{  \Gamma_{k}}}\mathrm{tan}\theta\big) \bm{I}}{\Phi^{-1}(\eta_{k}) }& \bm{\Theta}^{\frac{1}{2}}_{k,2}[\bm{u}_R,\bm{u}_I]\\
( \bm{\Theta}^{\frac{1}{2}}_{k,2}[\bm{u}_R,\bm{u}_I] )^T& \frac{\big(\overline{-\bm{a}_{k,2}}^T[\bm{u}_R,\bm{u}_I]+\sigma_n\sqrt{ \overline{  \Gamma_{k}}}\mathrm{tan}\theta\big)}{\Phi^{-1}(\eta_{k}) } \\
\end{smallmatrix}
\right] \succeq \bm{0}
\end{array}  
\right. 
\label{eq:Schur2}
\end{equation}  

Defining $\bm{U}=\bm{u}^H\bm{u}$, P2(imperfect-prob) is further transformed to

\begin{small}
\begin{equation}
\begin{split}
& P3~ (\mathrm{imperfect-prob}): \operatorname*{argmin}\limits_{\bm{u},\bm{t}} \dfrac{ \mathrm{Tr}(\bm{U})}{\alpha} + \sum_{n=1}^{N}\big(t_np_{on}+(1-t_n)p_{off} \big),\\
&~\mathrm{s.t~}(C1):  \mathrm{Tr}(\bm{UF}_n) \leq t_n p_{DA}, \forall n \in N,~ (C2),~(C3):  (\ref{eq:Schur1}),~(C4):  (\ref{eq:Schur2}), \forall k \in K,\\
& ~~~~~(C5):  \left[
\begin{matrix}
\bm{U}& \bm{u}\\
\bm{u}^T& 1
\end{matrix}
\right] \succeq 0, (C6):  \bm{U}  \succeq 0.
\label{eq:probabilistic_P3}
\end{split}
\end{equation}
\end{small}%
where $F_n=\mathrm{diag}( \underbrace{0 ...0}_{n-1},1,\underbrace{0, ...0}_{N-n})$ is an auxiliary diagonal matrix whose elements are zero except the $n$-th element, $\forall k \in K$. The optimization problem in (\ref{eq:probabilistic_P3}) is still non-convex due to the binary variables in $(C2):$ $t_n =\{0,1\}, \forall n \in N$, which can be relaxed to the following two constraints \cite{Derrick2016Power}

\begin{small}
\begin{equation}
\begin{split}
& (C2a):t_n=[0,1]  ,\forall n\in N,~(C2b):\sum_{n=1}^{N}t_n-\sum_{n=1}^{N}t_n^2 \leq 0.
\label{eq:equivalent binary}
\end{split}
\end{equation}
\end{small}%
where $(C2a)$ is the relaxed version of the original constraint, and $(C2b)$ confines the value of $t_n$ close to 0 or 1.  
Introducing a penalty factor $\varphi$, typically of large value, and moving $(C2b)$ into the objective function, the objective becomes into $\dfrac{ \mathrm{Tr}(\bm{U})}{\alpha} + \sum_{n=1}^{N}\big(t_np_{on}+(1-t_n)p_{off} \big) +\varphi \big(  \sum_{n=1}^{N}t_n-\sum_{n=1}^{N}t_n^2 \big)$, which shares the same optimal design policy and result with the original problem \cite{Derrick2016Power}. The last difficulty lies in the non-convex term $\varphi(\sum_{n=1}^{N}t_n-\sum_{n=1}^{N}t_n^2)$ in the objective term. It can be observed that $\sum_{n=1}^{N}t_n-\sum_{n=1}^{N}t_n^2$ is the difference of two convex functions w.r.t the variable $t_n$, and thus can be handled by successive convex approximation such that $\sum_{n=1}^{N}t_n-\sum_{n=1}^{N}t_n^2 \leq \sum_{n=1}^{N} t_n-\sum_{n=1}^{N}( t_n^{(i)})^2-2\sum_{n=1}^{N}  t_n^{(i)} (t_n-t_n^{(i)})$, where $t_n^{(i)}$ denotes the value of $t_n$ at the $i$-th iteration \cite{Nguyen2013Precoding}. Therefore, the optimization becomes into

\begin{small}
\begin{equation}
\begin{split}
& P4~ (\mathrm{imperfect-prob}):  \\
& \operatorname*{argmin}\limits_{\bm{u},~\bm{t}} \dfrac{ \mathrm{Tr}(\bm{U})}{\alpha} + \sum_{n=1}^{N}\big(t_np_{on}+(1-t_n)p_{off} \big)+\varphi \big(  \sum_{n=1}^{N} t_n-\sum_{n=1}^{N}(t_n^{(i)})^2-2\sum_{n=1}^{N}  t_n^{(i)} (t_n-t_n^{(i)})  \big),\\
&~\mathrm{s.t}~ (C1),~(C2a): t_n=[0,1],\forall n\in N, (C3)-(C6).
\label{eq:probabilistic_P4}
\end{split}
\end{equation}
\end{small}%
	
Now P4(imperfect-prob) is a standard semi-definite programming (SDP) problem, which can be readily solved by CVX. To tighten the approximation, we update the value of $t_n^{(i)}$ at each iteration. The successive convex approximation serves as the upper bound of the original problem, which is iteratively minimized and also lower bounded by the SINR and physical layer security constraints. Hence, the convergence of the algorithm is confirmed. Finally, the solver for the P4(imperfect-prob) in (\ref{eq:probabilistic_P4}) is summarized as follows.


\begin{algorithm}
	\caption{DA system with Imperfect CSI and probabilistic robust optimization (DA-imperfect-prob)}
	\label{alg:Algorithm1}
	\footnotesize
	\begin{algorithmic}[1]
		\renewcommand{\algorithmicrequire}{ \textbf{Input:}} 
		\renewcommand{\algorithmicensure}{ \textbf{Output:}} 
		\REQUIRE Probabilities thresholds $\eta_{d}$ and $\eta_{k}$. SINR requirements $\overline{\Gamma_{d}}$ and $\overline{\Gamma_{k}}$. 
		 Estimated CSI  $ \hat{\bm{h}_{d}}, ~ \hat{\bm{h}_{k}},~ \forall k \in K,$  and power consumption parameters $\alpha,~ p_{on},~ p_{off},~ p_{DA} $.
		\ENSURE
		Optimal precoding $\bm{u}$ and DA selection $\bm{t}$.
		\REPEAT
		\STATE Solve optimization problem P4(imperfect-prob) in (\ref{eq:probabilistic_P4}).
		\STATE $t_n^{(i)}=t_n, \forall ~n \in N$. 
		\STATE $i=i+1$.
		\UNTIL{Convergence}	
	\end{algorithmic}
\end{algorithm}

\textbf{\textit{Remark 1}}:
Now we revisit the inequalities in (\ref{eq:equivalent IR8}). Since the 2-norm term $ || \bm{\Theta}^{\frac{1}{2}}_{d,1} [\bm{u}_R,\bm{u}_I] ||_2 \geq 0$  always holds, the first inequality in (\ref{eq:equivalent IR8}) actually contains two inequalities (also applicable for the second inequality  in (\ref{eq:equivalent IR8}))

\begin{small}
	\begin{equation}   
	-\overline{\bm{a}_{d,1}}^T [\bm{u}_R,\bm{u}_I] \geq \sigma_n\sqrt{ \overline{  \Gamma_{d}}}\mathrm{tan}\theta,~  
	|| \bm{\Theta}^{\frac{1}{2}}_{d,1} [\bm{u}_R,\bm{u}_I] ||_2 \leq \frac{ -\overline{\bm{a}_{d,1}}^T [\bm{u}_R,\bm{u}_I] -\sigma_n\sqrt{ \overline{  \Gamma_{d}}}\mathrm{tan}\theta}{\Phi^{-1}(\eta_{d} )},
	\label{eq:property 1}
	\end{equation}  
\end{small}%
where the first inequality of (\ref{eq:property 1}) suggests that the magnitude of $\bm{u}$ should be large enough while the second inequality of (\ref{eq:property 1}) suggests that the magnitude of $\bm{u}$ should be smaller enough to make the (\ref{eq:property 1}) holds. Hence, given a high channel estimation error (the norm of matrix $ \bm{\Theta}^{\frac{1}{2}}_{d,1} $ is large), the value of $\eta_d$ or $\overline{\Gamma_d}$ needs to be properly reduced to make the optimization feasible.


\textbf{\textit{Remark 2}}:
Revisiting the constraint for the $k$-th Eve in (\ref{eq:equivalent_eve3}), we get

\begin{small}
	\begin{equation}  
	|| \bm{\Theta}^{\frac{1}{2}}_{k,1}[\bm{u}_R,\bm{u}_I] ||_2 \leq \frac{\sigma_n\sqrt{ \overline{  \Gamma_{k}}}\mathrm{tan}\theta-\overline{\bm{a}_{k,1}}^T[\bm{u}_R,\bm{u}_I]}{\Phi^{-1}(\eta_{k} )} ,~
	\overline{\bm{a}_{k,1}}^T[\bm{u}_R,\bm{u}_I] \leq \sigma_n\sqrt{ \overline{  \Gamma_{k}}}\mathrm{tan}\theta,  
	\label{eq:property eve}
	\end{equation}  
\end{small}%
where only upper bounds on the magnitude of $\bm{u}$ are imposed. As a result, the constraint itself always confines a feasible region.  



\subsection{Deterministic Robust Optimization}

In the previous subsection, we have solved the problem in a probabilistic manner, where the IR's QoS requirement and the physical layer security against Eves are issued by the chance constrained formulations. In this section, we handle the CSI uncertainties from a deterministic manner, where the IR's QoS requirement and the physical layer security towards the Eves are satisfied all the time with the infinite CSI uncertainties.
Define $\bm{\Delta}$ as the channel estimation uncertainties set, which contains all the possible CSI uncertainties and specifies an ellipsoidal uncertainty region for the estimated CSI \cite{Boshkovsha2015Practical}. 
\subsubsection{Problem Formulation}
To process the power minimization problem in terms of deterministic robust optimization, the formulation is given as 

\begin{small}
\begin{equation}
\begin{split}
& P5~ (\mathrm{imperfect-det}): \operatorname*{argmin}\limits_{\bm{w},\bm{z},\bm{t}}   \dfrac{||  \bm{w}+\bm{z}e^{-j\phi_{d}}  ||^2}{\alpha}    + \sum_{n=1}^{N}\big(t_np_{on}+(1-t_n)p_{off} \big),\\
&~~\mathrm{s.t~}(C7):0\leq |w_n+z_ne^{-j\phi_{d}}|^2\leq t_np_{DA}, \forall n\in N,~(C8):t_n=\{0,1\}  ,\forall n\in N,\\
& ~~~~~~~(C9):  \operatorname*{min}\limits_{\bm{e_{d}\in \Delta}} \Gamma_{d} \geq \overline{ \Gamma_{d} }, ~(C10): \operatorname*{max}\limits_{\bm{e_{k}\in \Delta}} \Gamma_{k} \leq \overline{ \Gamma_{k} }, k \in K ,  
\label{eq:deterministic_P1}
\end{split}
\end{equation}
\end{small}%
where $(C9)$ and $(C10)$ indicate deterministic SINR requirement for the IR and physical layer security constraints against Eves, such that the IR's worst-case SINR and Eves' best-case SINR as per the CSI error distribution obey the respective thresholds $\overline{\Gamma_d}$ and $\overline{\Gamma_k},$ $\forall k \in K.$

\subsubsection{Optimization Solution}

In line with the analysis in previous section and defining $\bm{w}+\bm{z}e^{-j\phi_{IR}}=\bm{u}$, constraint $(C9)$ is equivalent to the following two inequalities


\begin{small}
	\begin{equation}  
	\left\{  
	\begin{array}{lr}  
\operatorname*{min}\limits_{\bm{e_{d}\in \Delta}} [ \bm{e}_{I,d} -\bm{e}_{R,d}\mathrm{tan}\theta; \bm{e}_{R,d} +\bm{e}_{I,d}\mathrm{tan}\theta ]^T [\bm{u}_R;\bm{u}_I]+\varrho_{d,1} \leq 0,\\
\operatorname*{min}\limits_{\bm{e_{d}\in \Delta}} [ -\bm{e}_{I,d} -\bm{e}_{R,d}\mathrm{tan}\theta; -\bm{e}_{R,d} +\bm{e}_{I,d}\mathrm{tan}\theta ]^T [\bm{u}_R;\bm{u}_I]+\varrho_{d,2} \leq 0,
\end{array}  
\right.  
\label{eq:determinstic equivalent IR}
\end{equation}  
\end{small}%
where $\varrho_{d,1}=\sigma_n\sqrt{ \overline{  \Gamma_{d}}} \mathrm{tan}\theta+\hat{ \bm{h}_{I,d}} \bm{u}_R+ \hat{ \bm{h}_{R,d}} \bm{u}_I-  \hat{ \bm{h}_{R,d}} \bm{u}_R\mathrm{tan}\theta+ \hat{ \bm{h}_{I,d}} \bm{u}_I \mathrm{tan}\theta$, $\varrho_{d,2}=\sigma_n\sqrt{ \overline{  \Gamma_{d}}} \mathrm{tan}\theta-\hat{ \bm{h}_{I,d}} \bm{u}_R- \hat{ \bm{h}_{R,d}} \bm{u}_I-  \hat{ \bm{h}_{R,d}} \bm{u}_R\mathrm{tan}\theta+ \hat{ \bm{h}_{I,d}} \bm{u}_I \mathrm{tan}\theta$.
We now handle the first inequality of (\ref{eq:determinstic equivalent IR}). To facilitate the infinite CSI uncertainties, we transform it into a linear matrix inequality (LMI) using the following Lemma 1:




\textbf{\textit{Lemma 1}} (S-Procedure): Let a function $ f_m({\bm{x}})$, $m \in \{1,2\}$, be defined as

\begin{small}
\begin{equation}
\begin{split}
f_m({\bm{x}})=\bm{x}^H\bm{A}_m\bm{x}+2\Re \{   \bm{b}_m^H\bm{x}  \}+c_m
\label{eq:determinstic equivalent IR3}
\end{split}
\end{equation}
\end{small}%
where $\bm{A}_m \in \mathbb{H}^{N\times N}$, $\bm{b}_m \in \mathbb{C}^{N\times1}$ and $c_m \in \mathbb{R}$. The implication $ f_1({\bm{x}}) \Rightarrow  f_2({\bm{x}})$ holds if and only if there exists an $\lambda$ such that

\begin{small}
\begin{equation}
\begin{split}
\lambda \left[
\begin{matrix}
\bm{A}_1,& \bm{b}_1\\
\bm{b}_1^H,& c_1
\end{matrix}
\right]-
\left[
\begin{matrix}
\bm{A}_2,& \bm{b}_2\\
\bm{b}_2^H,& c_2
\end{matrix}
\right] \succeq \bm{0}.
\label{eq:determinstic equivalent IR4}
\end{split}
\end{equation}
\end{small}%

According to S-procedure, we need to guarantee the derivation in (\ref{eq:determinstic equivalent IR5}) hold as the premise for guaranteeing the first inequality in (\ref{eq:determinstic equivalent IR}) always hold, which can be given as 

\begin{small}
	\begin{equation}
	\begin{split}
	&[\sqrt{\bm{e}_{I,d} -\bm{e}_{R,d}\mathrm{tan}\theta};\sqrt{\bm{e}_{R,d} +\bm{e}_{I,d}\mathrm{tan}\theta  }] ^T\bm{I}_{2N}
	[\sqrt{\bm{e}_{I,d} -\bm{e}_{R,d}\mathrm{tan}\theta};\sqrt{\bm{e}_{R,d} +\bm{e}_{I,d}\mathrm{tan}\theta  }] - \sigma^2 \leq 0 \Rightarrow\\
	& [\sqrt{\bm{e}_{I,d} -\bm{e}_{R,d}\mathrm{tan}\theta};\sqrt{\bm{e}_{R,d} +\bm{e}_{I,d}\mathrm{tan}\theta  }] ^T \mathrm{diag} (\bm{u}_R;\bm{u}_I)
	[\sqrt{\bm{e}_{I,d} -\bm{e}_{R,d}\mathrm{tan}\theta};\sqrt{\bm{e}_{R,d} +\bm{e}_{I,d}\mathrm{tan}\theta  }]  + \varrho_{d,1} \leq 0.
	\label{eq:determinstic equivalent IR5}
	\end{split}
	\end{equation}
\end{small}

By applying S-procedure, the derivation in (\ref{eq:determinstic equivalent IR5}) holds if and only if there exists $\lambda$ such that the LMI constraint in (\ref{eq:determinstic equivalent IR7}) holds \footnote{Based on the properties of normal distributed variables, it is easy to obtain that the terms $\bm{e}_{I,d} -\bm{e}_{R,d}\mathrm{tan}\theta$ and $\bm{e}_{R,d} +\bm{e}_{I,d}\mathrm{tan}\theta$ share the same expectation and variance. Thus, the two terms also share the same uncertainty bound $\sigma^2$, and the uncertainty bound $\sigma^2$ is normally larger than the CSI variance $\sigma_{e}^2$ to include all the uncertainties.}


\begin{small}
\begin{equation}
\begin{split}
\left[
\begin{matrix}
\lambda_{d,1} \bm{I}_{2N}-\mathrm{diag}(\bm{u}_R;\bm{u}_I),& \bm{0}\\
\bm{0},&- \lambda_{d,1} \sigma^2-\varrho_{d,1}
\end{matrix}
\right]
\succeq \bm{0}.
\label{eq:determinstic equivalent IR7}
\end{split}
\end{equation}
\end{small}%
	
Now the first constraint in $(\ref{eq:determinstic equivalent IR})$ containing infinite possibilities is transformed into a deterministic LMI constraint. Similarly, the second inequality in (\ref{eq:determinstic equivalent IR}) can be transformed into



\begin{small}
\begin{equation}
\begin{split}
\left[
\begin{matrix}
\lambda_{d,2}\bm{I}_{2N}-\mathrm{diag}(\bm{u}_R;\bm{u}_I),& \bm{0}\\
\bm{0},& -\lambda_{d,2} \sigma^2-\varrho_{d,2}
\end{matrix}
\right]
\succeq \bm{0}.
\label{eq:determinstic equivalent IR9}
\end{split}
\end{equation}
\end{small}%

Now the constraint $(C9)$ containing infinite possibilities is transformed into two deterministic LMI inequalities in (\ref{eq:determinstic equivalent IR7}) and (\ref{eq:determinstic equivalent IR9}), respectively.
Now we handle the $k$-th Eve's constraint in $(C10)$, which can be re-written in the two following constraints



\begin{small}
\begin{equation}  
\left\{  
\begin{array}{lr}  
\operatorname*{max}\limits_{\bm{e_{k}\in \bm{\Delta}}} 
\lbrack \bm{e}_{I,k} + \bm{e}_{R,k} \mathrm{tan}\theta; \bm{e}_{R,k} -\bm{e}_{I,k} \mathrm{tan}\theta \rbrack^T \lbrack\bm{u}_R;\bm{u}_I \rbrack+\varrho_{k,1}  \leq 0 , &  \\  
\operatorname*{max}\limits_{\bm{e_{k}\in \bm{\Delta}}} 
\lbrack-\bm{e}_{I,k} +\bm{e}_{R,k} \mathrm{tan}\theta; -\bm{e}_{R,k} -\bm{e}_{I,k} \mathrm{tan}\theta \rbrack^T \lbrack\bm{u}_R;\bm{u}_I\rbrack+\varrho_{k,2}  \leq 0  & \\  
\end{array}  
\right.  
\label{eq:determinstic equivalent IR11}
\end{equation}
\end{small}%
where $\varrho_{k,1}=\hat{\bm{ h}_{I,k}} \bm{u}_{R}+\hat{\bm{ h}_{R,k}} \bm{u}_{R}  \mathrm{tan}\theta+\hat{\bm{ h}_{R,k}} \bm{u}_{I}-\hat{\bm{ h}_{I,k}} \bm{u}_{I}\mathrm{tan}\theta-\sigma_n\sqrt{ \overline{  \Gamma_{k}}}\mathrm{tan}\theta$, and $\varrho_{k,2}=-\hat{\bm{ h}_{I,k}} \bm{u}_{R}+\hat{\bm{ h}_{R,k}} \bm{u}_{R}  \mathrm{tan}\theta-\hat{\bm{ h}_{R,k}} \bm{u}_{I}-\hat{\bm{ h}_{I,k}} \bm{u}_{I}\mathrm{tan}\theta-\sigma_n\sqrt{ \overline{  \Gamma_{k}}}\mathrm{tan}\theta$. Based on S-procedure introduced above, (\ref{eq:determinstic equivalent IR11}) can be transformed into

\begin{small}
\begin{equation}  
\left\{  
\begin{array}{lr}  
\left[
\begin{matrix}
\lambda_{k,1}\bm{I}_{2N}-\mathrm{diag}(\bm{u}_R;\bm{u}_I),&  \bm{0} \\
\bm{0}^T& -\lambda_{k,1} \sigma^2-\varrho_{k,1} \\
\end{matrix}
\right], &  \\  
\\
\left[
\begin{matrix}
\lambda_{k,2} \bm{I}_{2N}-\mathrm{diag}(\bm{u}_R;\bm{u}_I),& \bm{0}\\
\bm{0}^T& -\lambda_{k,2} \sigma^2-\varrho_{k,2} \\
\end{matrix}
\right]
\end{array}  
\right. 
\label{eq:determinstic equivalent IR12}
\end{equation} 
\end{small}%

Now the non-convex constraints $(C9)$ and $(C10)$ are replaced by equivalent LMI constraints. 
Defining $\bm{U}=\bm{u}\bm{u}^H$, the problem can be transformed as

\begin{small}
\begin{equation}
\begin{split}
& P6~ (\mathrm{imperfect-det}): \operatorname*{argmin}\limits_{\bm{w},\bm{z},\bm{t}} \dfrac{ \mathrm{Tr}(\bm{U})}{\alpha} + \sum_{n=1}^{N}\big(t_np_{on}+(1-t_n)p_{off} \big),\\
&\mathrm{s.t}~(C7):  \mathrm{Tr}(\bm{UF}_n) \leq t_n p_{DA}, \forall n \in N,~(C8):t_n=\{0,1\}  ,\forall n\in N,\\
&~~~~~(C9): (\ref{eq:determinstic equivalent IR7})~ \mathrm{and}~(\ref{eq:determinstic equivalent IR9}),~(C10):  (\ref{eq:determinstic equivalent IR12}), \forall k \in K,~(C11):  \left[
\begin{matrix}
\bm{U}& \bm{u}\\
\bm{u}^T& 1
\end{matrix}
\right] \succeq 0,~(C12):  \bm{U} \succeq 0,\\
&~~~~~(C13): \lambda_{d,1} \geq 0,~(C14): \lambda_{d,2} \geq 0,~(C15): \lambda_{k,1} \geq 0,~(C16): \lambda_{k,2} \geq 0, \forall k \in K.
\label{eq:determinstic P2}
\end{split}
\end{equation}
\end{small}%

To handle the binary variable in $(C8)$, we relax it by a real number between [0,1] and add a penalty factor $\varphi$ in the objective function. Then the whole problem can be given as

\begin{small}
\begin{equation}
\begin{split}
& P7~ (\mathrm{imperfect-det}): \\
&\operatorname*{argmin}\limits_{\bm{w},\bm{z},\bm{t}} \dfrac{ \mathrm{Tr}(\bm{U})}{\alpha} + \sum_{n=1}^{N}\big(t_np_{on}+(1-t_n)p_{off} \big)+\varphi \big(  \sum_{n=1}^{N} t_n-\sum_{n=1}^{N}(t_n^{(i)})^2-2\sum_{n=1}^{N}  t_n^{(i)} (t_n-t_n^{(i)})  \big),\\
& \mathrm{s.t}~(C7),~(C8): t_n \in [0,1], (C9)-(C16).
\label{eq:imperfect determinstic P3}
\end{split}
\end{equation}
\end{small}%

Now the solver for the deterministic robust optimization problem is summarized as follows.

\begin{algorithm}
	\caption{DA systems with Imperfect CSI and deterministic robust optimization (DA-imperfect-det)}
	\label{alg:Algorithm2}
	\footnotesize
	\begin{algorithmic}[1]
		\renewcommand{\algorithmicrequire}{ \textbf{Input:}} 
		\renewcommand{\algorithmicensure}{ \textbf{Output:}} 
		\REQUIRE  SINR requirements $\overline{\Gamma_{d}}$ and $\overline{\Gamma_{k}}$. Estimated channel $ \hat{\bm{h}_{d}},  \hat{\bm{h}_{k}},~ \forall k \in K$. Power consumption parameters $\alpha, ~p_{on},~ p_{off},~p_{DA} $. 
		\ENSURE
		Optimal precoding $\bm{u}$ and DAs activation/deactivation $\bm{t}$.
		\REPEAT
		\STATE Solve optimization problem $P4$ in (\ref{eq:imperfect determinstic P3}).
		\STATE $t_n^{(i)}=t_n, \forall ~n \in N$. 
		\STATE $i=i+1$.
		\UNTIL{Convergence}	
	\end{algorithmic}
\end{algorithm}

\section{Power Efficient DA Selection and Precoding With Unknown  Eves' CSI}
In a number of practical scenarios, it is impossible to obtain Eves' CSI information. In this section, we investigate physical layer security issue when imperfect CSI is only available for the IR while the Eves' CSI is completely unknown. Again, we solve the problem in probabilistic and deterministic manners.

\subsection{Probabilistic Robust Optimization}
Since the IR' CSI can be imperfectly obtained, we can still guarantee the IR' SINR requirement by a probabilistic constraint. However, to address physical layer security in absence of any Eves' CSI, we can only set a minimum power level of AN.

\subsubsection{Problem Formulation}
To minimize the total power consumption subjected to the IR's SINR constraint and Eves' security constraints, the problem is formulated as 

\begin{small}
\begin{equation}
\begin{split}
& P8~ (\mathrm{unknown-prob}):\operatorname*{argmin}\limits_{\bm{w},~\bm{z},~ \bm{t}}   \dfrac{||  \bm{w}+\bm{z}e^{-j\phi_{d}}  ||^2}{\alpha}    + \sum_{n=1}^{N}\big(t_np_{on}+(1-t_n)p_{off} \big),\\
&~~\mathrm{s.t~}(C17):0\leq |w_n+z_ne^{-j\phi_{d}}|^2\leq t_np_{DA}, \forall n\in N,~(C18):t_n=\{0,1\}  ,\forall n\in N,\\
& ~~~~~~(C19): \mathrm{Pr}\{\ \Gamma_{d} \geq \overline{ \Gamma_{d} } |  \bm{e}_{d}   \} \geq \eta_{d},~(C20):||  \bm{z}  ||^2 \geq P_{AN},
\label{eq:unknown_P1}
\end{split}
\end{equation}
\end{small}%
where constraint $(C20)$ is imposed to guarantee the minimum power level of AN and is of importance when Eves' CSI is unknown \cite{Mukherjee2014Principles}. 

\subsubsection{Optimization Solution}
To solve the problem, we transform the probabilistic constraint $(C19)$ into equivalent LMI constraints as we did in the previous section. Define $\bm{u}=\bm{w}+\bm{z}e^{-j\phi_{d}}$, $U=\bm{u}^H\bm{u}$ and $Z=\bm{z}^H\bm{z}$. The problem is given by

 \begin{small}
\begin{equation}
\begin{split}
& P9 ~(\mathrm{unknown-prob}):\operatorname*{argmin}\limits_{\bm{w},\bm{z},\bm{t}} \dfrac{ \mathrm{Tr}(\bm{U})}{\alpha} + \sum_{n=1}^{N}\big(t_np_{on}+(1-t_n)p_{off} \big)\\
&~\mathrm{s.t~}(C17): \mathrm{Tr}(\bm{UF}_n) \leq t_n p_{DA}, \forall n \in N,~(C18).~(C19):(\ref{eq:Schur1}),~ (C20):  \mathrm{Tr}(\bm{Z}) \geq p_{AN},\\
&~~~~~(C21): \mathrm{Tr}(\bm{ZF}_n) \leq \mathrm{Tr}(\bm{UF}_n), ~(C22):\bm{w}+\bm{z}e^{-j\phi_{d}}=\bm{u},\\
& ~~~~~(C23):  \left[
\begin{matrix}
\bm{U}& \bm{u}\\
\bm{u}^T& 1
\end{matrix}
\right] \succeq 0,~(C24):  \left[
\begin{matrix}
\bm{Z}& \bm{z}\\
\bm{z}^T& 1
\end{matrix}
\right] \succeq 0,~(C25):  \bm{U} \succeq 0,~(C26):  \bm{Z} \succeq 0.
\label{eq:unknown pro 2}
\end{split}
\end{equation}
\end{small}%
where constraint $(C21)$ is imposed to guarantee that the AN generated on each DA is lower than the overall beamformer weight. Then we relax the binary variable $t_n, \forall k \in K$ $(C18)$ into a real value between [0,1] and add a penalty factor $\varphi$ into the objective function. The transformed problem is given as

\begin{small}
\begin{equation}
\begin{split}
& P10~ (\mathrm{unknown-prob}):\\
&\operatorname*{argmin}\limits_{\bm{w},\bm{z},\bm{t}} \dfrac{ \mathrm{Tr}(\bm{U})}{\alpha} + \sum_{n=1}^{N}\big(t_np_{on}+(1-t_n)p_{off} \big)+\varphi \big(  \sum_{n=1}^{N} t_n-\sum_{n=1}^{N}(t_n^{(i)})^2-2\sum_{n=1}^{N}  t_n^{(i)} (t_n-t_n^{(i)})  \big),\\
&~\mathrm{s.t}~(C17),~(C18):t_n=[0,1] ,\forall n\in N,~(C9)-(C26).
\label{eq:unknown pro 3}
\end{split}
\end{equation}
\end{small}%

By iteratively updating the value of DA selection vector $\bm{t}$, the optimization problem can be readily solved. The solver is summarized in Algorithm 3.

\begin{algorithm}
	\caption{DA systems with unknown Eves' CSI and probabilistic robust optimization (DA-unknown-prob)}
	\label{alg:Algorithm3}
	\footnotesize
	\begin{algorithmic}[1]
		\renewcommand{\algorithmicrequire}{ \textbf{Input:}} 
		\renewcommand{\algorithmicensure}{ \textbf{Output:}} 
		\REQUIRE Minimum AN power level $p_{AN}$. Probability threshold $\eta_d$. SINR requirement of the IR $\overline{\Gamma_d}$.		
		Estimated CSI of the IR  $ \hat{\bm{h}_{d}}$. Power consumption parameters, $\alpha,~ p_{on},~ p_{off},~p_{DA} $. 
		\ENSURE
		Optimal precoding $\bm{u}$ and DAs activation/deactivation $\bm{t}$.
		\REPEAT
		\STATE Solve optimization problem $P2$ in (\ref{eq:unknown pro 2}).
		\STATE $t_n^{(i)}=t_n, \forall ~n \in N$. 
		\STATE $i=i+1$.
		\UNTIL{Convergence}	
	\end{algorithmic}
\end{algorithm}

\subsection{Deterministic Robust Optimization}
Now we present the power efficient design in the deterministic manner, for the case when the Eves' CSI is unknown. As discussed above, the IR's SINR needs to be satisfied with all the CSI uncertainties. Again, we set a minimum power level of AN to address physical layer security issue.

\subsubsection{Problem Formulation}
The problem is formulated as 

\begin{small}
\begin{equation}
\begin{split}
& P11 ~(\mathrm{unknown-det}): \operatorname*{argmin}\limits_{\bm{w},\bm{z},\bm{t}}   \dfrac{||  \bm{w}+\bm{z}e^{-j\phi_{d}}  ||^2}{\alpha}    + \sum_{n=1}^{N}\big(t_np_{on}+(1-t_n)p_{off} \big),\\
&~\mathrm{s.t~}(C27):0\leq |w_n+z_ne^{-j\phi_{d}}|^2\leq t_np_{DA}, \forall n\in N,~(C28):t_n=\{0,1\}  ,\forall n\in N,\\
& ~~~~~(C29):  \operatorname*{min}\limits_{\bm{e_{d}\in \Delta}} \Gamma_{d} \geq \overline{ \Gamma_{d} },  ~(C30):||  \bm{z}  ||^2 \geq P_{AN},
\label{eq:unknown_P2}
\end{split}
\end{equation}
\end{small}%
where $(C29)$ denotes that the minimum possible SINR of the IR should be higher than the required $\overline{\Gamma_d}$ with all the CSI uncertainties. 

\subsubsection{Optimization Solution}
As discussed above, we first transform the constraint $(C29)$ that contains infinity probabilities into an equivalent LMI by applying S-procedure. Defining $ \bm{w}+\bm{z}e^{-j\phi_{d}}=\bm{u}$ as well as relaxing the binary variable $t_n, \forall n \in N$, the transformed problem is given as 

\begin{small}
\begin{equation}
\begin{split}
& P12~(\mathrm{unknown-det}):  \operatorname*{argmax}\limits_{\bm{w},\bm{z}, \bm{t}} \\
&\dfrac{ \mathrm{Tr}(\bm{U})}{\alpha} + \sum_{n=1}^{N}\big(t_np_{on}+(1-t_n)p_{off} \big)+\varphi \big(  \sum_{n=1}^{N} t_n-\sum_{n=1}^{N}(t_n^{(i)})^2-2\sum_{n=1}^{N}  t_n^{(i)} (t_n-t_n^{(i)})  \big),\\
&\mathrm{s.t}~(C27):  \mathrm{Tr}(\bm{UF}_n) \leq t_n p_{DA}, \forall n \in N,~(C28):t_n=\{0,1\}  ,\forall n\in N,~(C29): (\ref{eq:determinstic equivalent IR7})~ \mathrm{and}~(\ref{eq:determinstic equivalent IR9}),\\
&~~~~~(C30):  \mathrm{Tr}(\bm{Z}) \geq p_{AN},~(C31): \mathrm{Tr}(\bm{ZF}_n) \leq \mathrm{Tr}(\bm{UF}_n), ~(C32): \bm{w}+\bm{z}e^{-j\phi_{d}}=\bm{u},\\
&~~~~~(C33):  \left[
\begin{matrix}
\bm{U}& \bm{u}\\
\bm{u}^T& 1
\end{matrix}
\right] \succeq 0,~(C34):  \left[
\begin{matrix}
\bm{Z}& \bm{z}\\
\bm{z}^T& 1
\end{matrix}
\right] \succeq 0,~(C35):  \bm{U} \succeq 0,~(C36):  \bm{Z} \succeq 0,\\
& ~~~~~(C37): \lambda_{d1} \geq 0, (C38): \lambda_{d2} \geq 0.
\label{eq:determinstic P3}
\end{split}
\end{equation}
\end{small}%

Finally, the solver is given by Algorithm 4.
\begin{algorithm}
	\caption{DA systems with unknown Eves' CSI and deterministic robust optimization (DA-unknown-det)}
	\label{alg:Algorithm4}
	\footnotesize
	\begin{algorithmic}[1]
		\renewcommand{\algorithmicrequire}{ \textbf{Input:}} 
		\renewcommand{\algorithmicensure}{ \textbf{Output:}} 
		\REQUIRE CSI error bound $\sigma^2$. Minimum AN power level $p_{AN}$. SINR requirement of the IR $\overline{\Gamma_d}$.		
		Estimated CSI of the IR  $ \hat{\bm{h}_{d}}$. Power consumption parameters $\alpha, p_{on},~ p_{off},~p_{DA} $. 
		\ENSURE
		Optimal precoding $\bm{u}$ and DAs activation/deactivation $\bm{t}$.
		\REPEAT
		\STATE Solve optimization problem $P2$ in (\ref{eq:unknown pro 2}).
		\STATE $t_n^{(i)}=t_n, \forall ~n \in N$. 
		\STATE $i=i+1$.
		\UNTIL{Convergence}	
	\end{algorithmic}
\end{algorithm}

\section{COMPLEXITY ANALYSIS}

In this section we analytically examine the computational complexity of the proposed algorithms, and benchmark them against the closely related schemes in \cite{Ng2015Secure}, \cite{Khandaker2018Constructive}. For a fair comparison, the parameters in \cite{Ng2015Secure} \cite{Khandaker2018Constructive} have been modified to be consistent (the same number of users and antennas). It can be seen that the proposed algorithm runs in an iterative manner, and we have several LMI and linear inequalities to handle in each iteration. For the interior-point methods based solver, the overall complexity can be given as $\mathrm{ln}(\frac{1}{\epsilon})l_{i}\sqrt{c_{b}}(c_{f}+c_{g})$ \cite{Wang2014Outage}. Specifically, $\mathrm{ln}(\frac{1}{\epsilon})$ relates to the accuracy setup. $l_{i}$ represents the number of iterations for updating DA selection vector. $\sqrt{c_{b}}$ represents the barrier parameter measuring
the geometric complexity of the conic constraints. $c_{f}$ and $c_{g}$ represent the complexities cost on forming and factorization of $n\times n$ matrix of the linear system \footnote{By the interior-point methods based solver, a search direction is found by solving a system of linear equations in $n$ unknowns. $c_{f}$ is calculated as $c_{f}=n\sum_{j=1}^{P}k_j^3+n^2\sum_{j=1}^{P}k_j^2+n\sum_{j=P+1}^{m}k_j^2$, where $k_j$ presents the size of the $j$-th constraint. Specifically, the terms $n\sum_{j=1}^{P}k_j^3+n^2\sum_{j=1}^{P}k_j^2$ come from $P$ LMI constraints while the term $n\sum_{j=P+1}^{m}k_j^2$ comes from $m-P$ second order cone constraints in the problem formulation. $c_{g}$ is calculated as $c_{g}=n^3$ (Eq. (18), \cite{Wang2014Outage}). }.

We first investigate the algorithm's complexity when the IR and Eves' CSI is imperfectly obtained.
For the DA-imperfect-prob algorithm in (\ref{eq:probabilistic_P4}), it involves with $N$ LMI (trace) in $(C1)$ of size 1, $2N$ linear inequality in $(C2)$, 2 LMI in $(C3)$ of size $2N+1$, $2K$ LMI in $(C4)$ of size $2N+1$, 1 LMI in $(C5)$ of size $N+1$ and 1 LMI in $(C6)$ of size $N$. Hence, the proposed algorithm has $c_{b}=(9N+3)+2K(2N+1)$, $c_{f}=n( 3N+N^3+(N+1)^3+(2K+2)(2N+1)^3 +n^2(3N+N^2+(N+1)^2+(2+2K)(2N+1)^2 )$, and  $c_{g}=n^3$ (Eq. (18)-(19), \cite{Wang2014Outage}). 
For the deterministic optimization problem in (\ref{eq:imperfect determinstic P3}), it involves with $N$ LMI (trace) in $(C7)$ of size 1, $2N$ linear inequalities in $(C8)$, 2 LMI inequalities in $(C9)$ of size $2N+1$, $2K$ LMI inequalities in $(C10)$ of size $2N+1$,  1 LMI inequality in $(C11)$ of size $N+1$, 1 LMI inequality in $(C12)$ of size $N$, and $2+2K$ linear inequalities in constraints $(C13)-C(16)$. Hence, the proposed algorithm has $c_{b}=9N+2K+5+2K(2N+1)$, $c_{f}=n( 3N+2+2K+(2K+2)(N+1)^3+(N+1)^3+N^3) +n^2(3N+2+2K+(2K+2)(N+1)^2+(N+1)^2+N^2)$, and  $c_{g}=n^3$. 



Now we analyze the complexities with unknown Eves' CSI. For the probabilistic robust optimization in (\ref{eq:unknown pro 3}), it involves with $N$ LMI (trace) in $(C17)$ of size 1, $2N$ linear constraints in $(C18)$, 2 LMI (trace) in $(C19)$ of size $(2N+1)$, 1 LMI (trace) in $(C20)$ of size 1, $N$ LMI (trace) in $(C21)$ of size 1, $N$ linear constraints in $(C22)$, 1 LMI in $(C23)$ of size $N+1$, 1 LMI in $(C24)$ of size $N+1$, and 2 LMI in $(C25)$ and $(C26)$ of size N, respectively. Hence, the algorithm has $c_{b}=13N+5$, $c_{f}=n( 5N+2N^3+1+2(N+1)^3+2(2N+1)^3 )+n^2( 5N+2N^2+1+2(N+1)^2+2(2N+1)^2    )    $, and  $c_{g}=n^3$. 
For the deterministic robust optimization in (\ref{eq:determinstic P3}), it involves with $N$ LMI (trace) in $(C27)$ of size 1, $2N$ linear constraints in $(C28)$, 2 LMI (trace) in $(C29)$ of size $(2N+1)$, 1 LMI (trace) in $(C30)$ of size 1, $N$ LMI (trace) in $(C31)$ of size 1, $N$ linear constraints in $(C32)$, 1 LMI in $(C33)$ of size $N+1$, 1 LMI in $(C34)$ of size $N+1$, 2 LMI in $(C35)$ and $(C36)$ of size N, 2 linear constraints in $(C37)$ and $(C38)$. Hence, the algorithm has $c_{b}=13N+7$, $c_{f}=n( 5N+3+2N^3+2(N+1)^3+2(2N+1)^3 )+n^2( 5N+3+2N^2+2(N+1)^2+2(2N+1)^2   )    $, and  $c_{g}=n^3$. 



Finally, the overall complexities of the proposed algorithms are summarized in TABLE I. It is observed that the proposed algorithms have polynomial time computational complexity. Given that the number of iterations $l_{i}$ for updating DA section vector is low, the proposed algorithms can converge rapidly and this will be further demonstrated in the simulation section.  Besides, it can be observed that, compared to the probabilistic robust optimization, the deterministic robust optimization has $2+2K$ more linear inequalities in the first scenario and $2$ more linear inequalities in the second scenario. However, the additional linear inequalities have minimal impact on the complexities as they only linearly increases with the number of users.


\begin{table*}
	\begin{center}
		\label{tab:simsetup}
		\footnotesize
		\centerline{TABLE \uppercase\expandafter{\romannumeral 1.} Complexity analysis, with accuracy factor $\epsilon$}
		\begin{tabular}{|c|c|} 
			\hline
			DA-imperfect-prob& $C_{ip}= \mathrm{ln}(\frac{1}{\epsilon})l_{i}\sqrt{ (9N+3)+2K(2N+1)} [ n( 3N+N^3+(N+1)^3+(2+2K)(2N+1)^3) $\\
			& $+n^2(3N+N^2+(N+1)^2+(2+2K)(2N+1)^2)+n^3    ], \mathrm{where}~n=\mathcal{O}(N+N^2)$\\	
			\hline
			DA-imperfect-det& $C_{id}=\mathrm{ln}(\frac{1}{\epsilon})l_{i}\sqrt {9N+4K+5+4K2N}  [    n( 3N+2+2K+(2+2K)(2N+1)^3$\\
			                &$+(N+1)^3+N^3) +n^2(3N+2+2K+(2K+2)(N+1)^2+(N+1)^2+N^2)+n^3$ \\
			\hline
			DA-unknown-prob &  $C_{up}= \mathrm{ln}(\frac{1}{\epsilon})l_{i}\sqrt {13N+5} [n( 5N+2N^3+1+2(N+1)^3+2(2N+1)^3 )$ \\
			                  &  $+n^2( 5N+1+2N^2+2(N+1)^2+2(2N+1)^2    )  +n^3 ] $\\
			\hline
			DA-unknown-det &  $ C_{ud}=\mathrm{ln}(\frac{1}{\epsilon})l_{i}\sqrt {13N+7} [n( 5N+3+2N^3+2(N+1)^3+2(2N+1)^3 )$ \\
			&  $+n^2( 5N+3+2N^2+2(N+1)^2+2(2N+1)^2    )  +n^3 ] $\\
			\hline
			CA-no-AS \cite{Khandaker2018Constructive}  &  $ \mathrm{ln}(\frac{1}{\epsilon}) \sqrt {(4+4K)} [n(8N^2+8KN^2)+n^3]$ , where $n=\mathcal{O}(2N)$\\
			\hline
			DA-conv-AN \cite{Ng2015Secure}&  $\mathrm{ln}(\frac{1}{\epsilon})l_{ite}\sqrt{2NL} [( 1+2K+3L )(2NL)^3+ (2NL)^2(1+2K+3L)^2+(1+2K+3L)^3].$ \\       
			\hline
		\end{tabular}
	\end{center}
\end{table*}

\section{Simulation Results}

We present the simulated performance in this section. The central frequency is set to 2 GHz with 1 MHz bandwidth. The AWGN power spectral density is -174 dBm/Hz. A $100\times100$ m$^2$ square cell model is considered with $N=16$ DAs, where the multiple DAs are uniformly fixed \cite{Wei2018Energy} across the map. The number of the Eves $K=14$. 
The IR and Eves are randomly distributed across the map. To further highlight the advantages of the DA deployment,  in Figs. \ref{fig:PercentagevsPower20180831}, \ref{fig:KvsPower20180831} and \ref{fig:Probabilities20180817}, subsets of the IR and Eves are partially placed at the coverage edge.
DE of all PAs is set to $\alpha=40\%$. Power consumption parameters are set to $p_{on}= 500$ mW, $p_{off}= 50$ mW, and $p_{DA}= 1000$ mW, respectively. CSI error is set to $\sigma_{d}=\sigma_{k}=0.01$. The PL model in \cite{Maria2008Polarized} is adopted. Besides,  the CA MISO system without antenna selection (CA-no-AS) in \cite{Khandaker2018Constructive} and the DA system with conventional AN (DA-conv-AN) in \cite{Ng2015Secure}  are selected as benchmarks. 
We also apply the proposed algorithms to CA systems, which are referred as CA-imperfect-prob,  CA-imperfect-det, CA-unknown-prob and CA-unknown-det respectively in simulation, and the multiple antennas of CA systems are co-located in the map center.

\subsection{Proposed Algorithms vs Benchmarks}

\begin{figure}
	\centering
	\includegraphics[width=3.7 in]{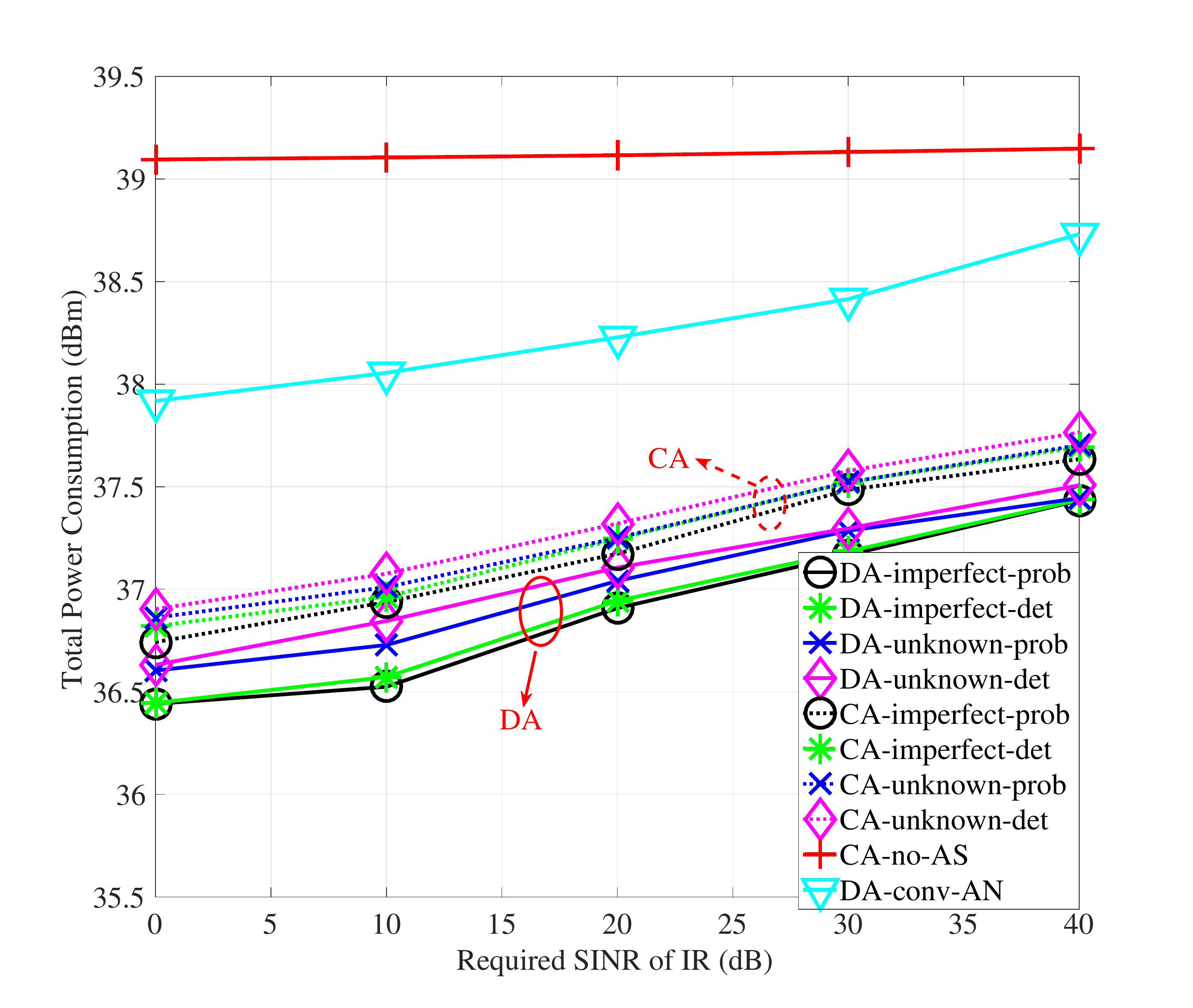}
	\caption{Impact of the IR's SINR requirement  $\overline{\Gamma_{d}}$ on the total power consumption, where $\overline{\Gamma_{k}}=-10$ dB, $\eta_d=\eta_k=0.95$, and $p_{AN}=25$ dBm.}
	\label{fig:IRSINRvsPower20180831}
\end{figure}

Fig. \ref{fig:IRSINRvsPower20180831} shows the impact of the IR's SINR requirement $ \overline{\Gamma_{d}}$ on the total power consumption with a random IR and Eve deployment.  
Firstly, it can be seen that the proposed algorithms outperform the two benchmarks, namely CA-no-AS and DA-conv-AN. It is because the proposed algorithms benefit from DAs' activation/deactivation, the DAs far from the users may be deactivated for saving power. Hence, the proposed algorithms achieve a user-centric and on-demand network structure with higher degree of freedom over the CA-no-AS algorithm in \cite{Khandaker2018Constructive}. Besides, our proposed algorithms rotate AN to make it constructive to the IR even with imperfect CSI, while AN is treated as an undesired element at the IR by conventional DA-conv-AN in \cite{Ng2015Secure}.
Secondly, benefiting from geometric distribution of the DAs, the distances between the DAs and users are shortened and the DA systems can always find near antennas to serve the users. With the  alleviated PL, DA can outperform its counterpart CA in terms of power efficient transmission.
Thirdly, increasing the SINR threshold $\overline{\Gamma_{d}}$ leads to higher power consumption, as more antennas become activated and dissipate higher circuit and transmission power. 
Also, the power consumption of the CA-no-AS increases slowly as the increased transmission power is overwhelmed by its circuit power consumption.
Fourthly, when the Eves' CSI in completely unknown, more power is dissipated compared to the scenario that the Eves' CSI is imperfectly obtained. It is because to address physical layer security, a minimum AN power level $p_{AN}$ is required, which is not efficient compared to the scenario that Eves' CSI is imperfectly obtained.

\begin{figure}
	\centering
	\includegraphics[width=3.7 in]{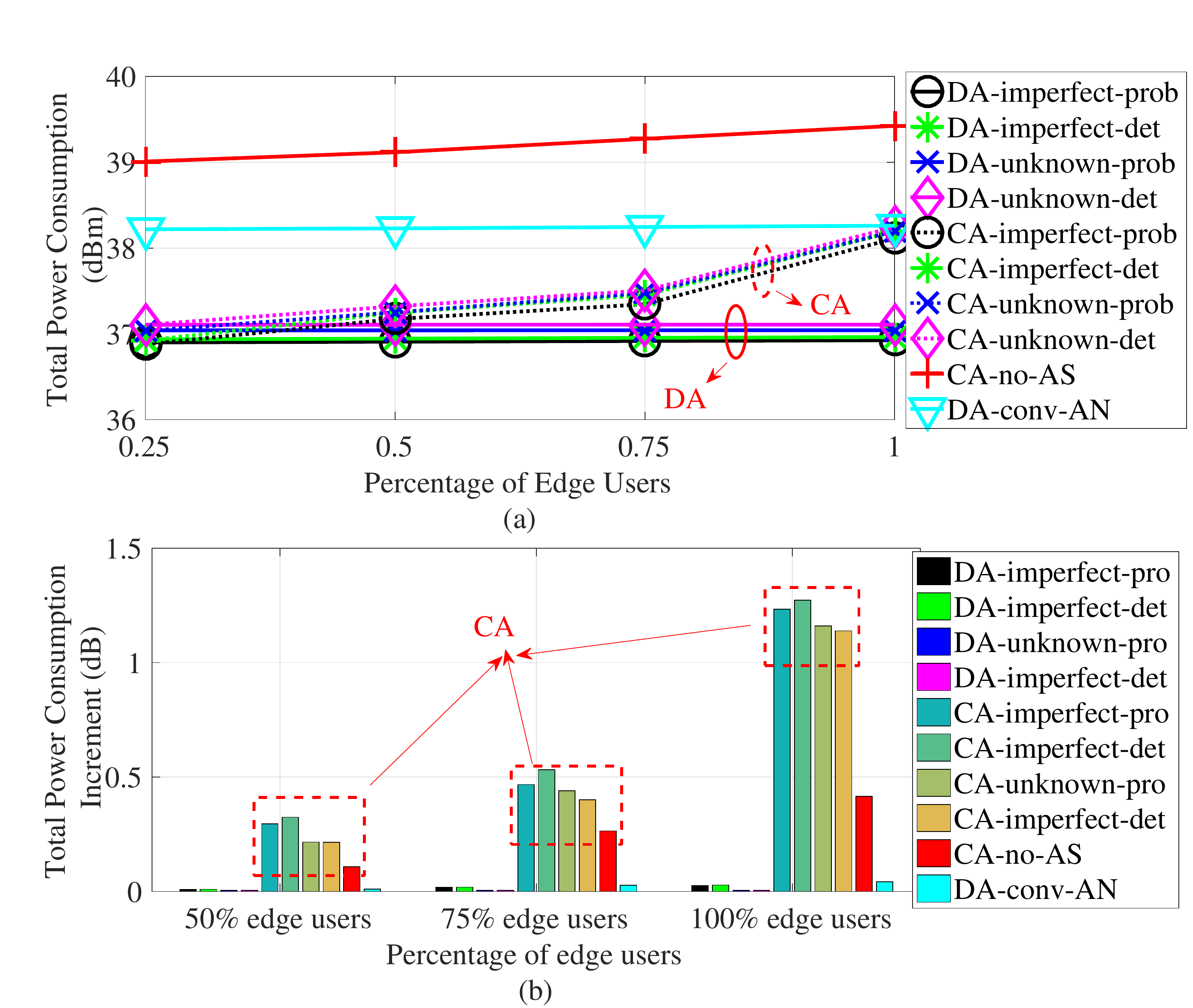}
	\caption{ (a) Impact of the percentage of edge on total power consumption.  (b) Power consumption increment (dB) of DA and CA systems with different percentages of edge users, where $\overline{\Gamma_{d}}=20$ dB, $\overline{\Gamma_{k}}=-10$ dB, $\eta_{d}=\eta_{k}=0.95$, and $p_{AN}=25$ dBm.}
	\label{fig:PercentagevsPower20180831}
\end{figure}

Fig.  \ref{fig:PercentagevsPower20180831} (a) shows the total power consumption with different percentages of edge users. It demonstrates that with more users at edge area, the power consumption of the DA systems almost remains unchanged regardless of the users' positions. It is because the geographically positioned DAs effectively extend the network coverage, and proposed algorithms always activate near DAs for serving the users.  By contrast, the total power consumption of the CA systems keeps increasing when more users move to edge area. This is because the CA systems have to activate more antennas and allocate higher transmission power to serve remote users, which inevitably further improve the total power consumption of CA systems. 
On the other hand, Fig.  \ref{fig:PercentagevsPower20180831} (b) shows the power consumption increment with different percentages of edge users, benchmarked by the power consumption with 25\% edge users. It is obvious that the power consumption of the CA systems  increases significantly with more edge users, and up to 1.3 dB increment of power consumption is achieved when all the users are located in the edge area. Besides, it is worthy noting that the power consumption increment of the CA-no-AS algorithm is lower than the other antenna selection-enabled CA systems, since the fully-activated antennas of the CA-no-AS have dominated the power consumption, which makes the increment less significant.

\begin{figure}
	\centering
	\includegraphics[width=3.7 in]{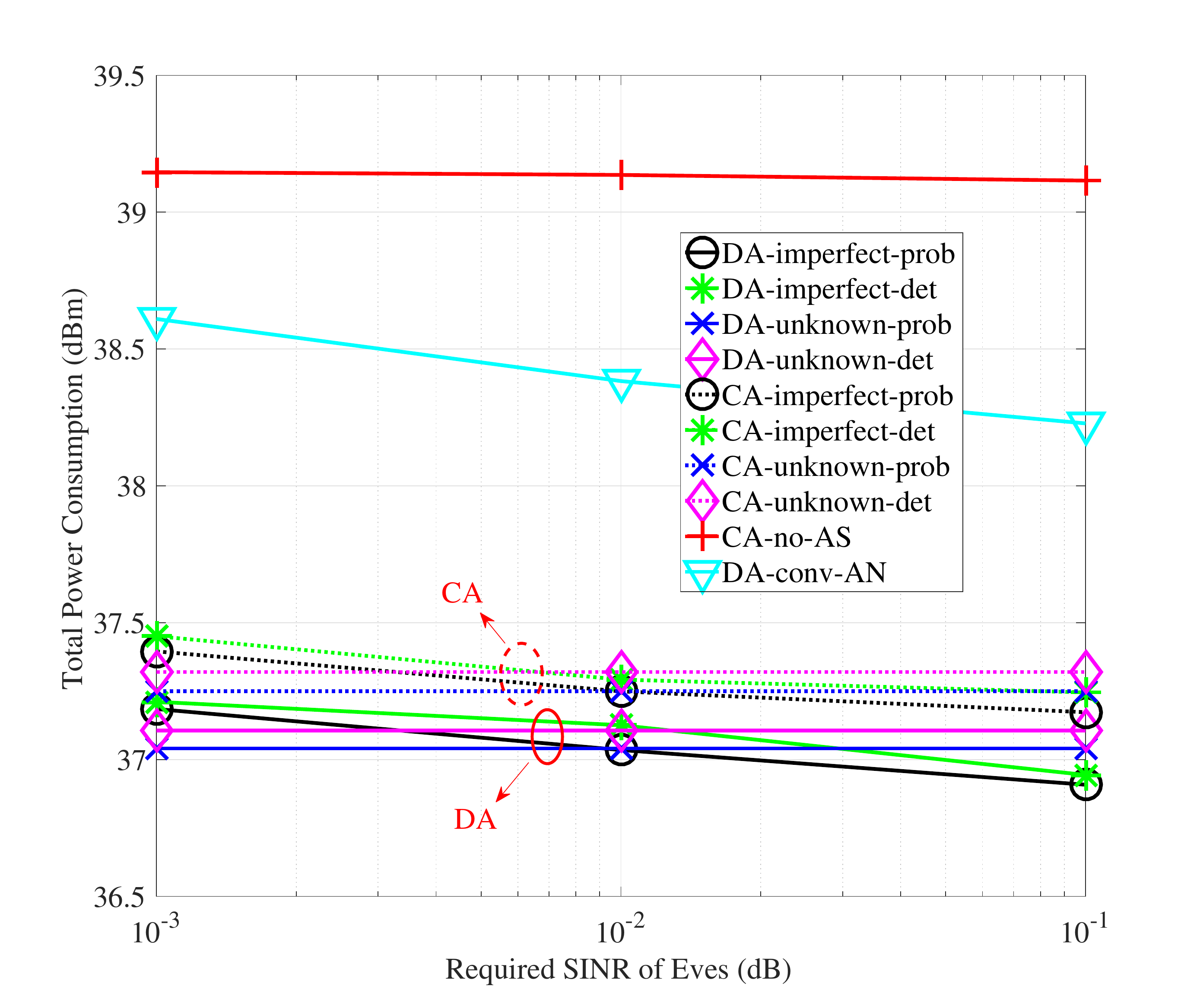}
	\caption{Power consumption of DA and CA systems with different SINR requirement against the Eves, where $\overline{\Gamma_{d}}=20$ dB, $\eta_{d}=\eta_{k}=0.95$, and $p_{AN}=25$ dBm.}
	\label{fig:EdgeEvesSINRvsPower20180831}
\end{figure}

Fig. \ref{fig:EdgeEvesSINRvsPower20180831} shows the impact of the physical layer security requirement $\overline{\Gamma_{k}}$ on the total power consumption. Firstly, a stringent physical layer security constraint, such as $\overline{\Gamma_{k}}=10^{-3}$, leads to higher power consumption compared to a loose physical layer security constraint. It is because with a stringent constraint, higher power level of AN is needed to make the Eves' SINR lower than the requirement $\overline{\Gamma_{k}}$.
Secondly, since the proposed algorithms efficiently utilize AN as a constructive element, the power consumption of the proposed algorithms maintain low and the physical layer security against the Eves can be simultaneously addressed. 
Thirdly, the DA systems always consume less power compared to the CA counterparts, benefiting from the user-centric structure. 
Fourthly, when the Eves' CSI is unknown at transmitter, the power consumption is independent with the value of $\overline{\Gamma_{k}}$. Hence, to address a stringent security requirement, one can properly increase the minimum power level of AN when Eves' CSI is unknown.

\begin{figure}
	\centering
	\includegraphics[width=3.7 in]{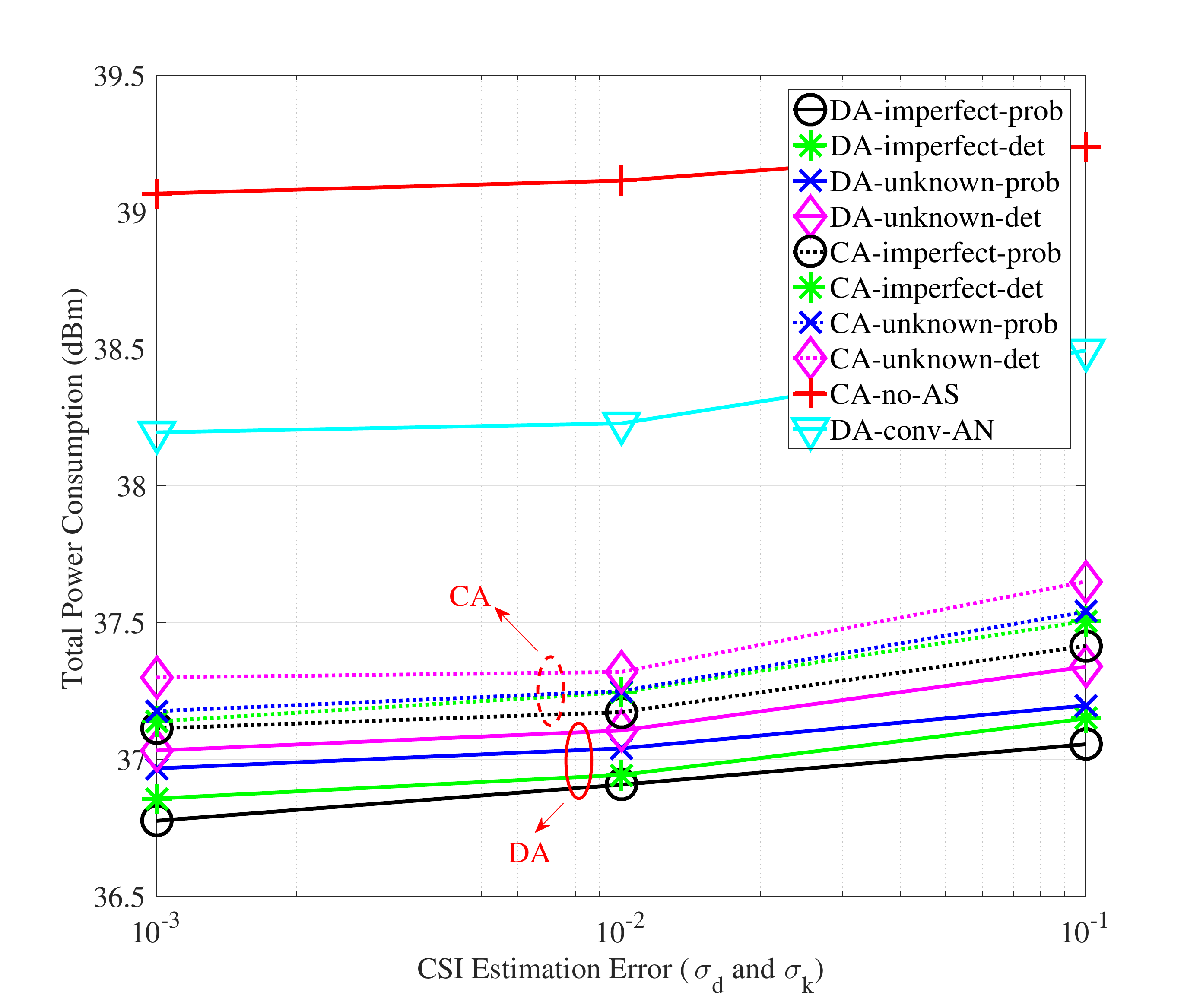}
	\caption{Impact of the CSI estimation error $\sigma_d$ and $\sigma_k$ on the total power consumption, where $\overline{\Gamma_{d}}=20$ dB, $\overline{\Gamma_{k}}=-10$ dB $\eta_d=\eta_k=0.95$, and $p_{AN}=25$ dBm. }
	\label{fig:ErrorvsPower20180905}
\end{figure}

Fig. \ref{fig:ErrorvsPower20180905} shows how the total power consumption is affected by the CSI estimation error. As can be seen, the power consumption of all the algorithms increases with a higher CSI error. In particular, for the probabilistic manner optimization, a higher CSI estimation error increases the norm of $\bm{\Theta}^{\frac{1}{2}}_{d,1}$ and thus the left hand of the first inequality in Eq. (\ref{eq:property 1}) also increases, as suggested by Remark 1. As a result, the amplitude of the precoder (also the transmission power) needs to be properly improved to make the optimization feasible, resulting in a increased total power.
On the other hand, the deterministic manner robust optimization needs to keep the positive semi-definite characteristic for the matrices in Eq. (\ref{eq:determinstic equivalent IR7}), (\ref{eq:determinstic equivalent IR9}) and (\ref{eq:determinstic equivalent IR12}). This mathematically requires that all the leading principal minors in the matrices to be non-negative.  Hence, with a higher CSI uncertainty, the amplitude of precoder (also the transmission power) needs to be property increased. As a result, the total power consumption of all the deterministic optimization is increased. A similar upwards trend can be also observed by the CA-no-AS and DA-conv-AN, which process the CSI uncertainties based on the deterministic manner as well. 

\begin{figure}
	\centering
	\includegraphics[width=3.7 in]{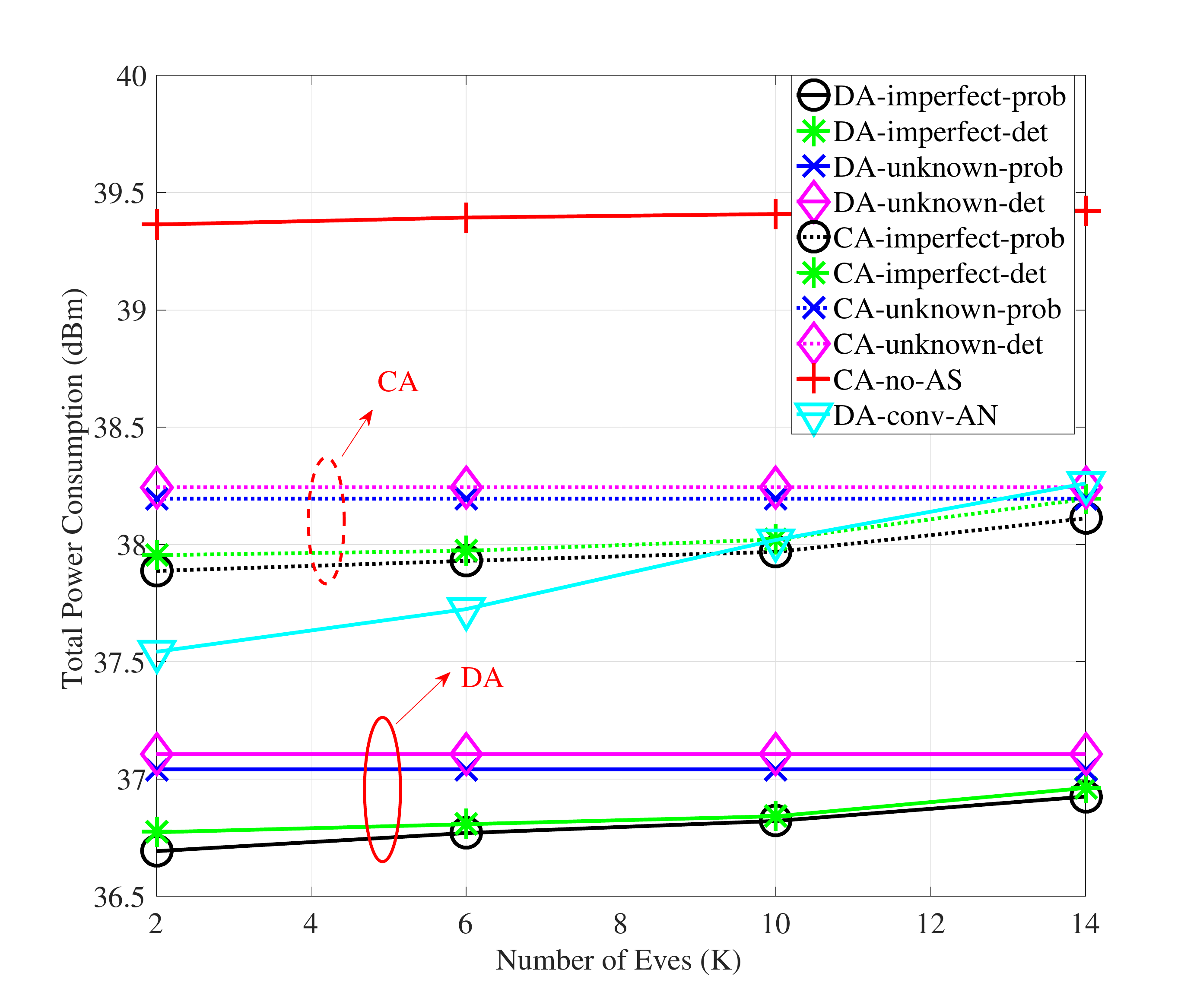}
	\caption{Impact of number of the Eves on total power consumption, where $\overline{\Gamma_{d}}=20$ dB, $\overline{\Gamma_{k}}=-10$ dB, $\eta_{d}=\eta_{k}=0.95$, and $p_{AN}=25$ dBm. }
	\label{fig:KvsPower20180831}
\end{figure}
\begin{figure}
	\centering
	\includegraphics[width=3.7 in]{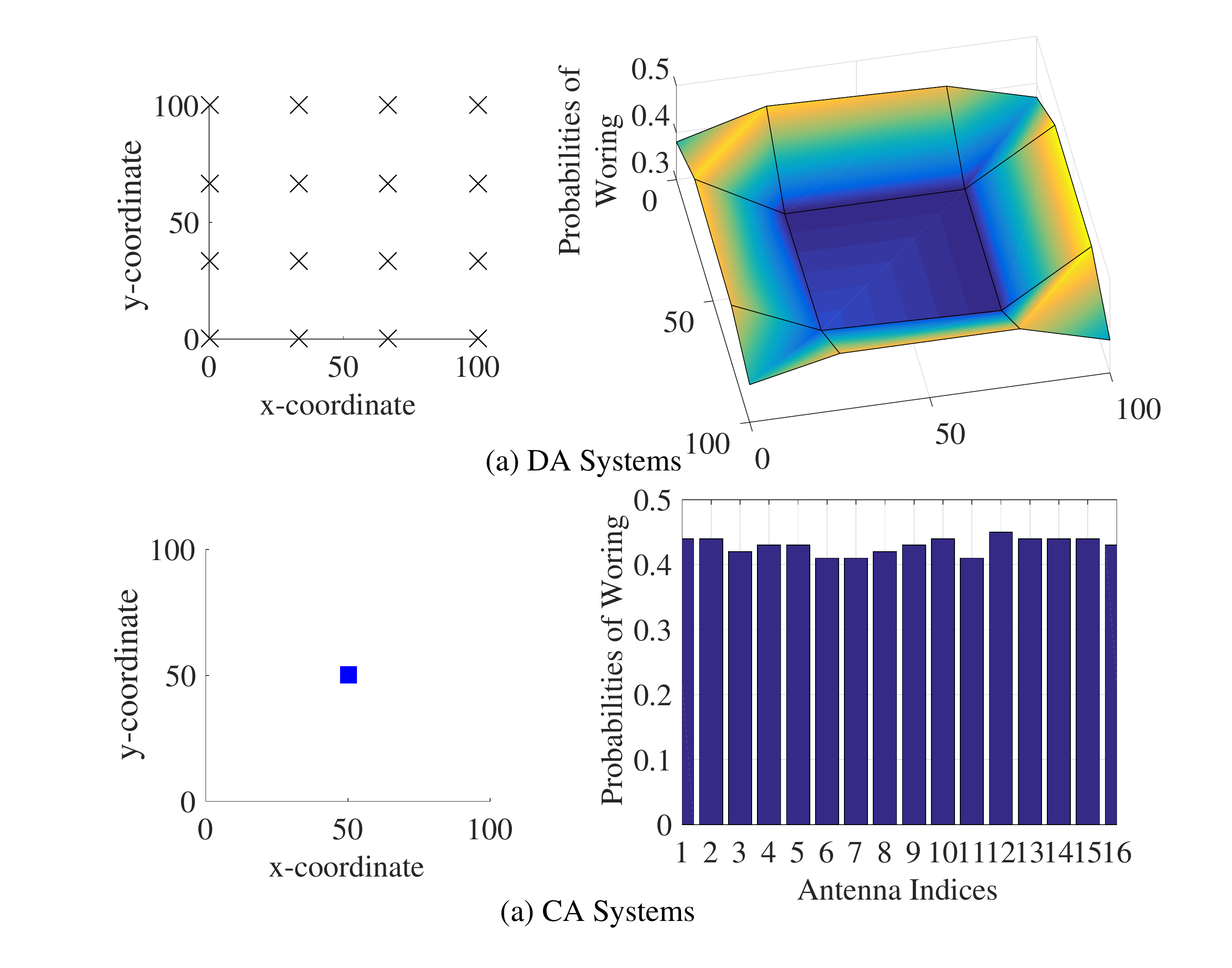}
	\caption{(a) Antenna deployment in DA systems, where antennas are geographically distributed. (b) Impact of users' positions on antennas' modes in DA systems. (c) Antenna deployment in CA systems, where antennas are co-located in the map center. (d)  Impact of users' positions on antennas' modes in CA systems. $\overline{\Gamma_{d}}=20$ dB and $\overline{\Gamma_{k}}=-10$ dB.  $\eta_{d}=\eta_{k}=0.95$, and $p_{AN}=25$ dBm. }
	\label{fig:Probabilities20180817}
\end{figure}

Now, in the above Figs. \ref{fig:IRSINRvsPower20180831}-\ref{fig:ErrorvsPower20180905}, we have demonstrated that the proposed algorithms outperform the benchmarks. Importantly, it is clear from the above results that the proposed algorithms are most beneficial for edge users. Hence, in Fig. \ref{fig:KvsPower20180831} and \ref{fig:Probabilities20180817}, we hereforth focus our attention on the scenario where all the IR and Eves are placed at the coverage edge to further highlight the advantages of DA deployment.

Fig. \ref{fig:KvsPower20180831}  shows how the total power consumption is affected by the number of Eves. Firstly, it can be seen that the DA deployment always outperforms its CA counterpart regardless of the number of the Eves. For the antenna selection-enabled CA systems, they almost dissipate the same power compared to the DA-conv-AN benchmark. Although artificial noise can be utilized by the antenna selection-enabled CA systems, the systems have to allocate higher transmission power for compensating severe PL, which is exactly avoided by the DA-conv-AN benchmark. As a result, the advantage of utilizing artificial noise vanishes in the antenna selection-enabled CA systems, especially in the scenario that all the IR and Eves are located at edge area.
Secondly, higher power for generating AN is required to maintain the physical layer security requirements when the number of the Eves increases, and thus higher total power consumption is led. Applying the concept of constructive AN yet pushing the IR's received symbols to the constructive region, the weight of precoder $\bm{w}$ can be interestingly reduced. As a result, the total power consumption of the proposed algorithms increase slowly with more Eves. By contrast, the two benchmarks dissipate more total power with the increased number of the Eves, either hindered by the high circuit power consumption at the multiple antennas (CA-no AS) or underutilized AN at the IR (DA-conv-AN). Especially for the DA-conv-AN scheme, since artificial noise is treated as a harmful element at the IR, its power consumption increases significantly compared to other schemes.
Thirdly, when the Eves' CSI is completely unaware by the system, the total power consumption remains unchanged with the increased number of Eves. However, one may preset a higher power level of $p_{AN}$ to address the physical layer security against the Eves, at the cost of high power consumption. 

Fig.  \ref{fig:Probabilities20180817} shows how antennas' modes (activation or deactivation) are affected by the users. For illustration, we only take the DA-imperfect-prob algorithm as an example due to the space constraints.
It can be seen from Figs. \ref{fig:Probabilities20180817}(a, b) that by the proposed algorithms, those DAs close to the users have higher probabilities of working while the central DAs far from the users have higher probabilities of being deactivated to save power. This is originated that in terms of power efficient design, letting those DAs far from the users transmit signal is not power efficient. 
By contrast, as can be seen in Fig. \ref{fig:Probabilities20180817}(c, d), the antennas in CA systems have the same probabilities of being activated or deactivated. It is because the centralized antennas have similar PL to the users, and their working modes are insensitive to users' positions. 

\subsection{CI Regions and Convergence Behavior}

\begin{figure}
	\centering
	\includegraphics[width=3.7 in]{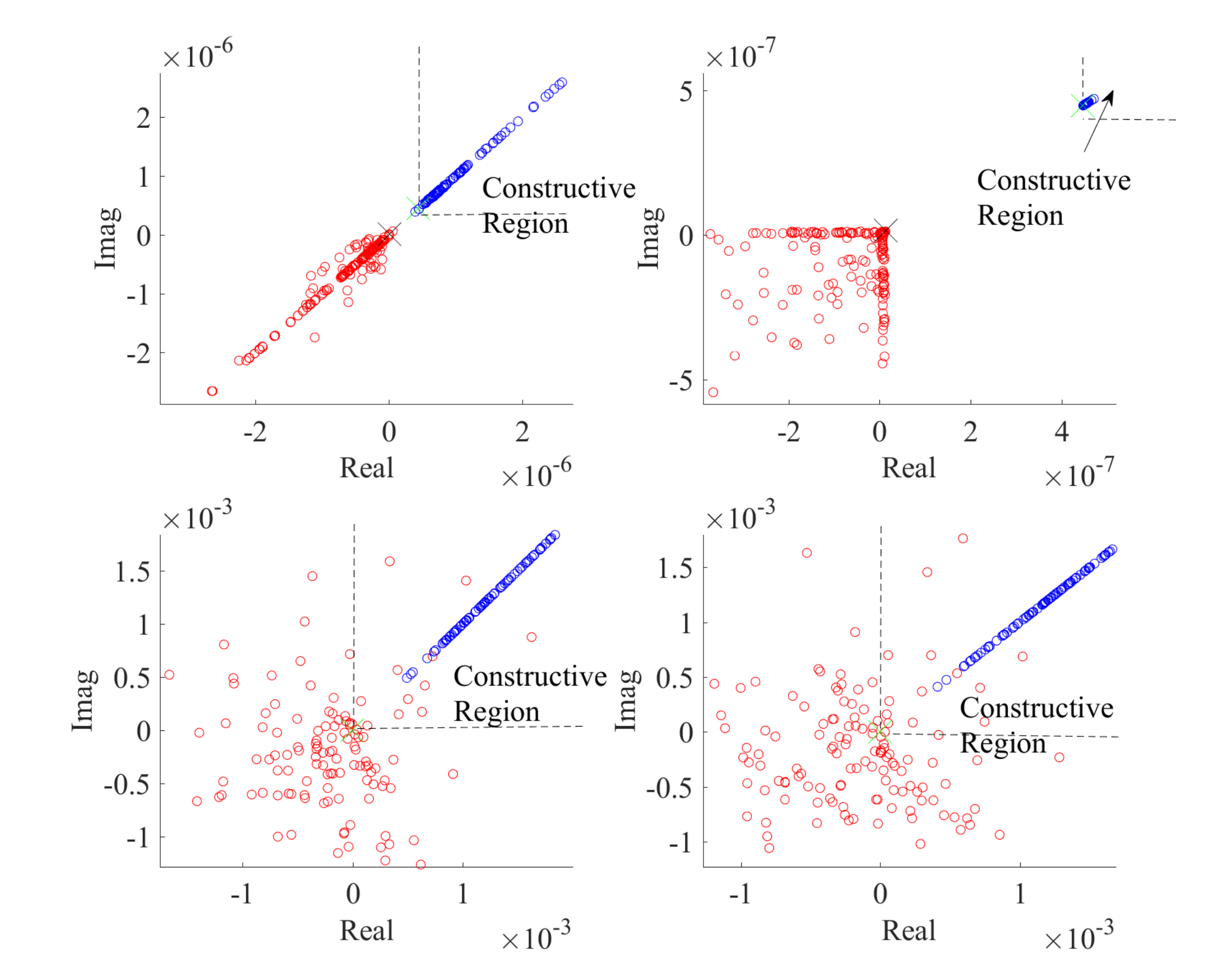}
	\caption{ The probabilities of the received symbols of the IR and Eves falling into constructive and destructive regions by the (a) DA-imperfect-prob, (b) DA-imperfect-det, (c) DA-unknown-prob, (d) DA-unknown-det algorithms, where red dots denote the received symbols of the Eves and blue dots denote the received symbols of the IR. $\overline{\Gamma_{d}}=20$ dB. $\overline{\Gamma_{k}}=-10$ dB, $\forall k \in K$. $\eta_{d}=\eta_{k}=0.95$ and $p_{AN}=25$ dBm. }
	\label{fig:Region20180723}
\end{figure}

For the DA-imperfect-prob algorithm in Fig. \ref{fig:Region20180723}(a), it can be seen that in most cases the received symbols of the IR locate in the constructive region while those of the Eves fall in destructive region as wanted, satisfying the preset outage probabilities $ \eta_{d}$ and $\eta_{k}, \forall k \in K$. This also verifies that even with DA selection mechanism to save power, the proposed algorithm still successfully keeps AN constructive for the IR whereas destructive for the Eves.
It is because the algorithm statistically guarantees the IR's SINR higher than $\overline{\Gamma_{d}}$ and the Eves' SINR lower than the security threshold $\overline{\Gamma_{k}}$. 
Fig. \ref{fig:Region20180723}(b) shows the probabilities under the DA-imperfect-det algorithm. According to our analysis in section III-B, the IR' SINR and Eves's security constraints should be satisfied with all the CSI uncertainties. This is verified by the simulation results that the received symbols of the IR always locate in the constructive region while those of the Eves locate in the destructive region all the time, which is essentially different from the probabilistic optimization that allows proper outage occurs. However, as discussed in Fig. 3, this is achieved at the cost of high power consumption.
Differently, Fig.  \ref{fig:Region20180723}(c) and (d) show the probabilities when the Eves' CSI is unknown by the system. It can be observed that both probabilistic and deterministic robust optimization can locate the received symbols of the IR in the constructive region, which means the IR's SINR requirement is readily satisfied and AN is also efficiently utilized. Besides, by setting the power level of AN higher than a threshold $p_{AN}$, the SINR of the Eves can be controlled lower than a threshold $\overline{\Gamma_k}$ in most cases, which addresses physical layer security against the Eves.

\begin{figure}
	\centering
	\includegraphics[width=3.7 in]{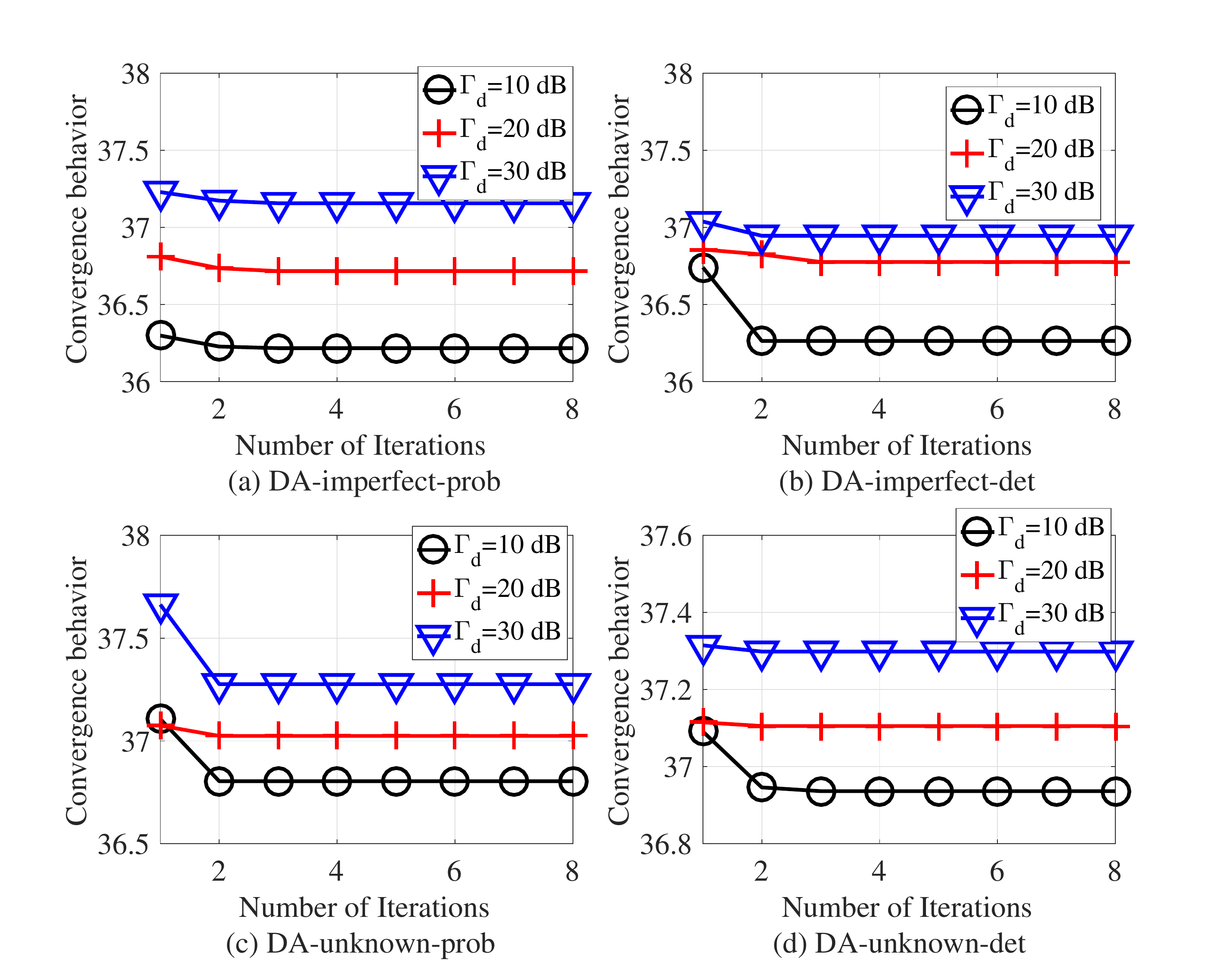}
	\caption{Iteration behavior of the proposed algorithms, where $\overline{\Gamma_{k}}=-10$ dB, $\forall k \in K.$ $\eta_{d}=\eta_{k}=0.95$, and $p_{AN}=25$ dBm.}
	\label{fig:Convergence20180723}
\end{figure}

Fig. \ref{fig:Convergence20180723} shows the convergence behavior of the proposed algorithms for updating antenna selection vector $\bm{t}$. It can be observed that the proposed algorithms are readily converged after 5 iterations, proving the low complexities of the algorithms.

\section{CONCLUSIONS}

We have investigated power minimization problem under physical layer security constraints for DA systems.
Targeting at the two practical scenarios, DA selection vector and precoding have been jointly optimized in terms of probabilistic and deterministic robust optimizations.
By the chance-constrained formulations as well as the deterministic formulations in the two robust optimizations, the proposed algorithms are able to satisfy the IR's QoS and simultaneously address physical layer security against the Eves. 
Our simulation results have showed that the proposed algorithms consume much lower power compared to the CA-no-AS and DA-conv-AN benchmarks, and low complexities have been confirmed by our computational analysis and simulation.
Furthermore, a flexible and user-centric network structure is featured benefiting from antenna selection mechanism in DA systems, and the power consumption is maintained at low level regardless of the users' positions compared to the CA counterpart.

\bibliographystyle{IEEEbib}
\bibliography{avrix20181212}

\end{document}